\documentclass[fleqn,usenatbib]{mnras}

\usepackage{newtxtext,newtxmath}
\usepackage[T1]{fontenc}
\usepackage{makecell}
\usepackage{comment}
\usepackage{hyperref}
\usepackage{subcaption}
\DeclareRobustCommand{\VAN}[3]{#2}
\let\VANthebibliography\thebibliography
\def\thebibliography{\DeclareRobustCommand{\VAN}[3]{##3}\VANthebibliography}

\usepackage{graphicx}	% Including figure files
\usepackage{amsmath}	% Advanced maths commands
\usepackage[nolist]{acronym}
\usepackage{xcolor}
\begin{acronym}

\acro{AGN}{Active Galactic Nuclei}
\acro{ASKAP}{Australian Square Kilometre Array Pathfinder}
\acro{CSIRO}{Commonwealth Scientific and Industrial Research Organisation}
\acro{DES}{Dark Energy Survey}
\acro{EMU}{Evolutionary Map of the Universe}
\acro{ICM}{intracluster medium}
\acro{BCG}{brightest cluster galaxy}
\acro{CARTA}{Cube Analysis and Rendering Tool for Astronomy}
\acro{RMS}{Root Mean Squared}
\acro{PSF}{point spread function}
\acro{RM}{rotation measure}
\acro{SNR}{signal-to-noise ratio}
\acro{FWHM}{full width half maximum}
\end{acronym}

\newcommand{\ghost}{PKS~2130--538}
\newcommand{\D}{$^\circ$}
\newcommand{\hour}{$^{\mathrm{h}}$}
\newcommand{\minute}{$^{\mathrm{m}}$}
\newcommand{\second}{$^{\mathrm{s}}$}

\title[the Dancing Ghosts]{MeerKAT view of the Dancing Ghosts --- Peculiar Galaxy Pair PKS~2130--538 in Abell~3785}

\author[V.Velovi\'{c} et al.]{Velibor Velovi\'{c}$^{1}$\thanks{E-mail: v.velovic@westernsydney.edu.au}, % \orcid{http://orcid.org/0000-0002-0416-3267}
William D. Cotton$^{2,3}$, % http://orcid.org/0000-0001-7363-6489
Miroslav D. Filipovi\'{c}$^{1}$, % http://orcid.org/0000-0002-4990-9288
Ray P. Norris$^{1,4}$, % http://orcid.org/0000-0002-4597-1906
Luke Barnes$^{1}$, % http://orcid.org/0000-0002-0016-9485
\newauthor 
James J. Condon${^2}$\\ % http://orcid.org/0000-0003-4724-1939
\\
$^{1}$School of Science, Western Sydney University, Locked Bag 1797, Penrith South DC, NSW 2751, Australia\\
$^{2}$National Radio Astronomy Observatory, 520 Edgemont Road, Charlottesville, VA 22903, USA \\
$^{3}$South African Radio Astronomy Observatory (SARAO), 2 Fir Street, Black River Park, Observatory, Cape Town 7925, South Africa \\
$^{4}$CSIRO Space and Astronomy, Australia Telescope National Facility, PO Box 76, Epping NSW 1710, Australia\\
}

\date{Accepted XXX. Received YYY; in original form ZZZ}

\pubyear{2022}

\begin{document}
\label{firstpage}
\pagerange{\pageref{firstpage}--\pageref{lastpage}}

\maketitle

% Abstract of the paper
\begin{abstract}
We present MeerKAT L--band (886--1682~MHz) observations of the extended radio structure of the peculiar galaxy pair \ghost\, known as the ``Dancing Ghosts''. The complex of bending and possibly interacting jets and lobes originate from two Active Galactic Nuclei hosts in the Abell~3785 galaxy cluster, one of which is the brightest cluster galaxy. The radio properties of the \ghost --- flux density, spectral index and polarization --- are typical for large, bent-tail galaxies. We also investigate a number of thin extended low surface brightness filaments originating from the lobes. Southeast from the Dancing Ghosts, we detect a region of low surface brightness emission that has no clear origin. While it could originate from the Abell~3785 radio halo, we investigate the possibility that it is associated with the two \ghost\ hosts. We find no evidence of interaction between the two \ghost\ hosts.

\end{abstract}
% Select between one and six entries from the list of approved keywords.
% Don't make up new ones.

%%%%%%%%%%%%%%%%%%%%%%%%%%%%%%%%%%%%%%%%%%%%%%%%%%

%%%%%%%%%%%%%%%%% BODY OF PAPER %%%%%%%%%%%%%%%%%%

\section{Introduction}
\label{sec:Intro}

The new generation of radio telescopes, including MeerKAT and \ac{ASKAP}, is discovering new objects and imaging them in greater detail than ever before \citep{MeerKAT_Heywood, EMU_Norris, ORCs, LMC_ORC}. Their wide-area coverage, in combination with the high sensitivity and spatial sampling, enables us to detect low surface brightness features and objects. 

One such complex object is the peculiar galaxy pair \ghost\ \citep[named as the Dancing Ghosts by][, see their Fig.~21]{EMU_Norris}. This extended radio source has been previously detected in high-resolution radio continuum surveys \citep{Ekers,Schilizzi_McAdam,Jones_McAdam} and reported as two radio galaxies (G4Jy 1704, 1705) by \citet{white_gleam_2,white_gleam_1}.

%%%%%%%%%%%%%%%%%%%%%%%%%%%%%% FIGURE 1 - DES %%%%%%%%%%%%%%%%%%%%%%%%%%%%%%
\begin{figure*}
\centering
\includegraphics[width=\textwidth,angle=0]{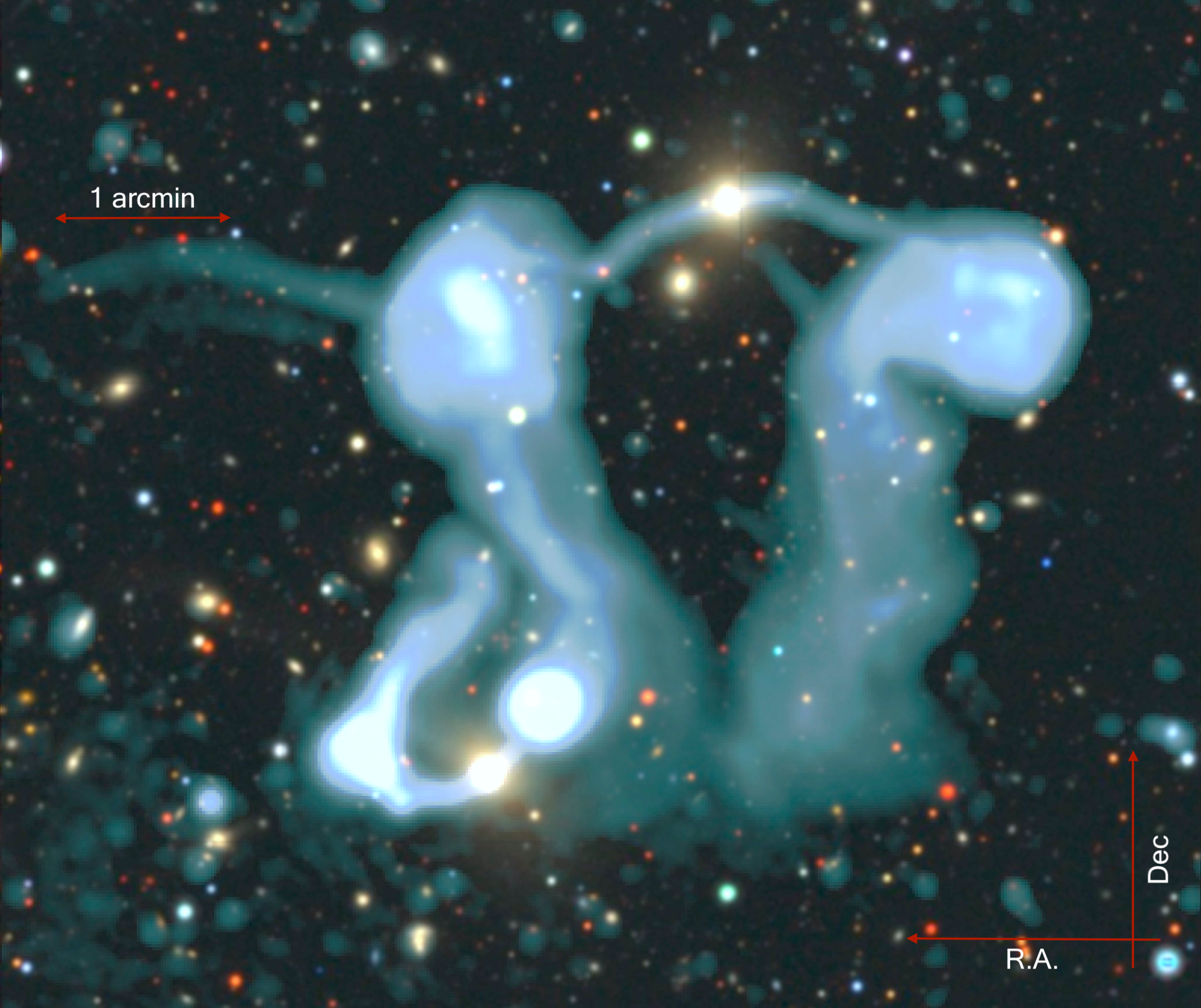}
    \caption{
    The total intensity image of \ghost, observed by MeerKAT at $\nu =1283$~MHz, split into three contrast levels and superimposed over the optical image obtained from the \ac{DES} legacy survey. To present the radio filamentary structure of the Dancing Ghosts, we adjusted linear contrast levels and applied a logarithmic colour scale. We also applied different linear colourmaps to enhance the filamentary and diffuse structure of the Dancing Ghosts. For all quantitative information refer to Figs.~\ref{figure:large_field}, \ref{figure:i_img_legend}, \ref{figure:tot_int} and \ref{figure:comparison}.
    }

\label{fig:DES_ghosts}
\end{figure*}
%%%%%%%%%%%%%%%%%%%%%%%%%%%%%%%%%%%%%%%%%%%%%%%%%%%%%%%%%%

Based on the \ac{ASKAP} Pilot Survey, \citet{EMU_Norris} argued that \ghost\ consists of two galaxies in the galaxy cluster Abell~3785, with possibly interacting radio lobes. The northern `Host~1', also identified as WISEA J213406.70-533418.7, sits in the centre of the northern radio bridge between two lobes at a photometric redshift of $z=0.078$ (with a comoving distance of $D_c \approx 328$~Mpc). `Host~2', identified as WISEA J213417.69-533811.1, is located in the southeast at a photometric redshift of $z=0.076$ ($D_c\approx 320$ Mpc) \citep{abell_members}. We adopt a flat $\Lambda$CDM cosmological model with $\textrm{H}_0= 70$~km s$^{-1}$ Mpc$^{-1}$, $\Omega_m=0.3$, $\Omega_\Lambda = 0.7$. At the redshift of Abell~3785 ($z = 0.0773$), we have $1''= 1.464$~kpc.

\citet[][see their Fig.~21]{EMU_Norris} indicated that the radio spectral index properties of \ghost\ are consistent with a bent-tail galaxy. At the position of Host~1, there is a flat-spectrum region \mbox{($\alpha \approx 0$)}\footnote{Radio spectral index $\alpha$ is defined as $S_{\nu}$~$\propto$~$\nu^\alpha$, where $S_\nu$ is the flux density at frequency $\nu$.} that connects to the relatively flat-spectrum jets ($\alpha \approx -0.4/-0.5$). These connect to large bright regions with spectral index $\alpha \approx$ --0.6 to --0.7, which steepens sharply as the lobes and tails change direction to the south to at least ($\alpha \approx -1.5$). Examination of the southern host shows a sudden change in the spectral index. Similar to the northern Host~1, it has a flat spectral index core and steeper spectral index lobe and tail structures. 

\citet[][]{EMU_Norris} were uncertain as to whether there is any interaction between the two hosts and their jets. They also noted a thin stream of emission extends from the north-eastern lobe with a median spectral index of $\alpha \approx -2.1$, which is not expected in existing radio galaxy models.

This paper is organised as follows: in Section~\ref{sec:data} we present data used in this study, while in Section~\ref{sec:results} we analyse MeerKAT observations of Abell~3785, followed by a detailed analysis of \ghost\ galaxy pair in the following Subsections~\ref{subsec:ghost}. In Section~\ref{sec:discussion} we discuss the results and possible interaction between the two hosts. Finally, we present our conclusions in Section~\ref{sec:conclusion}. Throughout the paper, we strictly use coordinates of the  equinox and epoch J2000.

\section{Data}
\label{sec:data}

South Africa's MeerKAT radio telescope is a precursor to the Square Kilometre Array (SKA). It consists of 64 antennas, each 13.5-m in diameter \citep{meerKAT}.  Observations of \ghost\ were conducted on 28$^{\rm th}$~November~2021 in L-band ($\nu=1283$~MHz; bandwidth of $800$~MHz) in 4k mode, for 10 hours including calibration. We used 61 of the 64 antennas.

For calibration, we used the pipeline of \citet{t_mauch}, followed by polarization calibration.
We used {\sc J1939--6342} as a delay, bandpass and leakage calibrator, while {\sc J2214--3835 }was used for complex gain and a leakage calibrator. Finally, we calibrated the residual X-Y phase difference using {\sc 3C138} and {\sc 3C286}.
The final spectrum, the total intensity at the reference frequency and spectral index, in pixels with adequate \ac{SNR}, was determined by a weighted least squares fit of the pixel values in the subband images. Pixels without adequate \ac{SNR} used the weighted average in frequency.

Final images were produced and cleaned using the Obit/{\sc MFImage} \citep{mfimage_cotton} using the robust weighting of $-1.5$, with 5~per~cent fractional bandwidth sub-bands for Stokes~I and 1~per~cent fractional bandwidths for Stokes Q and U with joint Q/U deconvolution. 1 and 5~per~cent fractional bandpass means that each subband had a bandwidth of 1 and 5~per~cent of the central frequency. All 14 channels with 5~per~cent fractional bandwidth subbands are given in Table~\ref{tab:channels}.
We applied frequency-dependent tapering to equalize the spatial resolution across the bandpass. 
The resulting image has a synthesized beam size of $7.5\times7.1$~arcsec$^2$ at P.A.$=0.82$\D\ and the local \ac{RMS} noise $1\sigma = 5.5\; \mu$Jy~beam$^{-1}$.

\begin{table}
\begin{center}
    \caption{Channelization and frequency details of 5~per~cent fractional bandwidth subbands used to produce the final images.}
    \begin{tabular}{ c c c c} 
    \hline
       \textbf{Channel} & \textbf{Low (MHz)} & \textbf{Center (MHz)} & \textbf{High (MHz)} \\ 
    \hline
        $1$   & $ 886.3 $  & $ 908.0 $  & $ 929.8 $  \\
        $2$   & $ 930.6 $  & $ 952.3 $  & $ 974.0 $  \\
        $3$   & $ 974.9 $  & $ 996.6 $  & $ 1018.4 $ \\
        $4$   & $ 1019.2 $ & $ 1043.5 $ & $ 1067.7 $ \\
        $5$   & $ 1068.5 $ & $ 1092.8 $ & $ 1117.0 $ \\
        $6$   & $ 1117.9 $ & $ 1144.6 $ & $ 1171.4 $ \\
        $7$   & $ 1172.2 $ & $ 1198.9 $ & $ 1225.7 $ \\
        $8$   & $ 1226.5 $ & $ 1255.8 $ & $ 1285.0 $ \\
        $9$   & $ 1285.9 $ & $ 1317.2 $ & $ 1348.6 $ \\
        $10$  & $ 1349.4 $ & $ 1381.2 $ & $ 1412.9 $ \\
        $11$  & $ 1413.8 $ & $ 1448.0 $ & $ 1482.3 $ \\
        $12$  & $ 1483.2 $ & $ 1519.9 $ & $ 1556.7 $ \\
        $13$  & $ 1557.6 $ & $ 1593.9 $ & $ 1630.3 $ \\
        $14$  & $ 1631.1 $ & $ 1656.2 $ & $ 1681.3 $ \\
    \hline
    \hline
    \end{tabular}
    \label{tab:channels}
\end{center}
\end{table}

\begin{figure*}
\includegraphics[width=\textwidth]{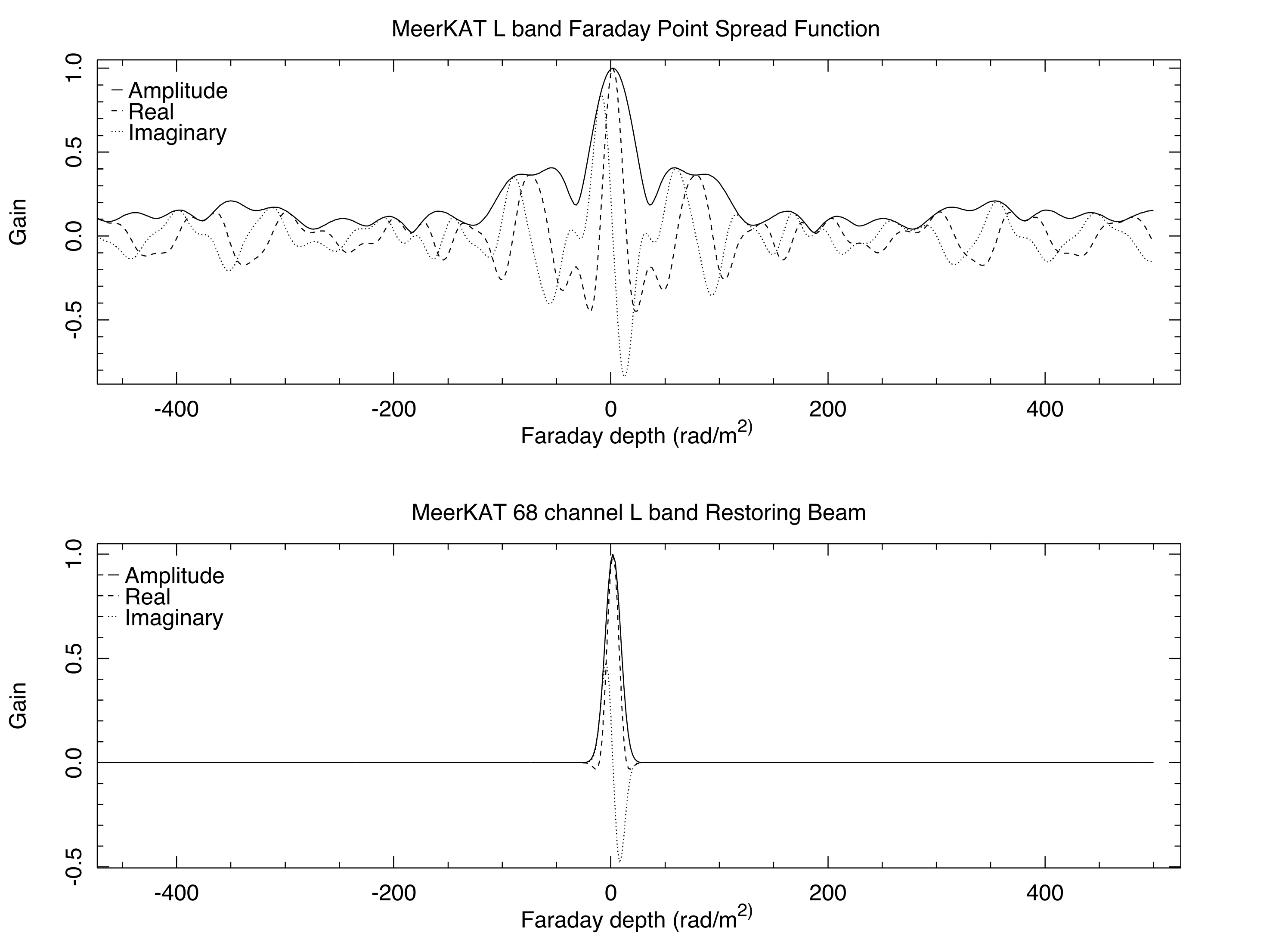}
\caption{{\em Top panel}: the complex ``dirty'' Faraday point spread function for MeerKAT L band data. The amplitude, real and imaginary parts are shown as solid, dashed and dotted lines. {\em Bottom panel}: the restoring function used for the deconvolved Faraday spectra. The width of the amplitude is that of the real component in the upper panel.}
\label{MKRMBeam}
\end{figure*}

The wide bandwidth of the MeerKAT L band data allows a Faraday synthesis analysis following that described in Rudnick \& Cotton (in prep.). This analysis uses a reference wavelength $\lambda_0 = 0$ which results in the complex Faraday \ac{PSF}) given in the upper panel of Fig.~\ref{MKRMBeam}. The high sidelobes from the spectral regions being blanked due to RFI require a complex deconvolution (CLEAN) to recover the Faraday spectrum. 
The \ac{FWHM} of the restoring function is $16$~rad~m$^{-2}$ (RMS$=6$~rad~m$^{-2}$). The Faraday analysis used $2$~rad~m$^{-2}$ elements covering the range $\pm500$~rad~m$^{-2}$ and corrected for an assumed spectral index of $\alpha = -0.7$. The Faraday synthesis \citep{RM} was performed using {\sc Obit} task {\sc RMSyn}.
Following Rudnick \& Cotton (in press), we use a restoring function for the width of the inner real lobe of the ``dirty'' \ac{PSF} as is shown in the lower panel of Fig.~\ref{MKRMBeam}. 

%%%% FIG 2 %%%%
%%%% FIG 2 %%%%

In addition to MeerKAT data, we also used \ac{ASKAP} data from \ac{EMU} survey described in \citet{EMU_Norris} and optical data from the DESI Legacy Imaging Surveys \citep{legacy}.

\section{Results}
\label{sec:results}

\begin{figure*}
\centering
\includegraphics[width=\textwidth,angle=0]{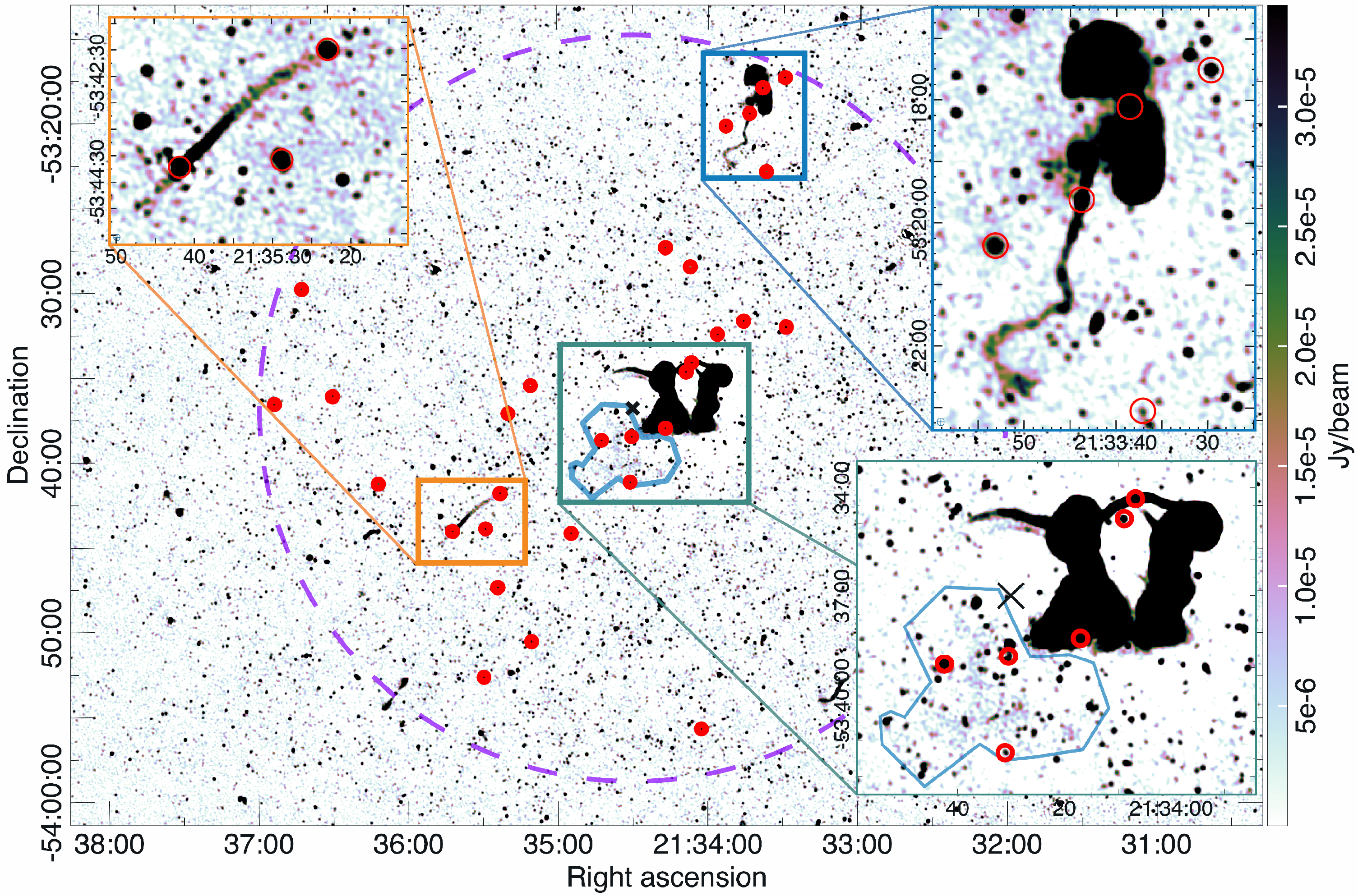}
\caption{MeerKAT image of the Dancing Ghosts field and the galaxy cluster Abell~3785. The dashed purple circle with a 22~arcmin radius marks the border of Abell~3785, with the centre of the cluster marked with a black ''X'' in the green square. Thirty radio continuum members of the Abell~3785 galaxy cluster are marked with red circles. The inset images are at the same intensity scale (colour bar) as the main image. The blue polygon marks the low surface brightness emission and three radio point sources belonging to the Abell~3785 galaxy cluster, just south of \ghost. The synthesized beam size (\ac{FWHM} of the PSF) is $7.5\times7.1$~arcsec$^2$ at P.A.$=0.82$~degrees and the \ac{RMS} is $\sigma = 5.5$\,$\mu$Jy~beam$^{-1}$.}
\label{figure:large_field}
\end{figure*}

\begin{figure*}
\centering
\includegraphics[width=\textwidth,angle=0]{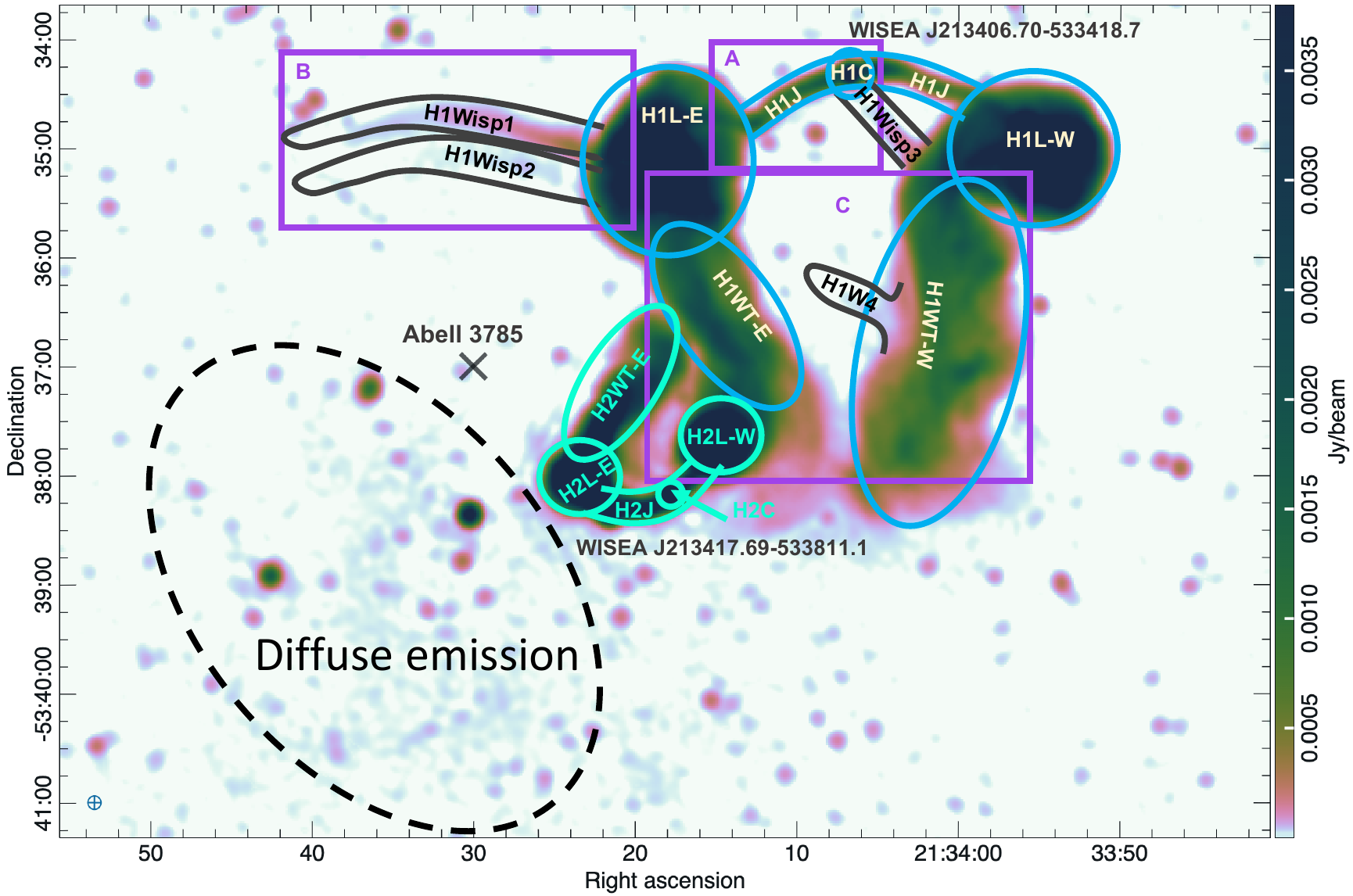}
    \caption{Total intensity image of \ghost\ obtained with MeerKAT at nominal frequency of $\nu=1283$~MHz. We used a logarithmic function to enhance low surface brightness. Legend: Host~1 (H1): H1C - core, H1J--Jets, H1L(W, E) -- lobes, H1Wisp(1,2,3,4) -- wisps, H1WT(W, E)--Wide Tails; Host~2 (H2): H2C -- Core; 8) H2L(W, E)-lobes 9) H2WT-E wide tail 10) dashed ellipse show the diffuse emission related to the source. The "X" marks the centre of Abell~3785. In the bottom left corner we show the synthesized beam size of $7.5\times7.1$~arcsec$^2$ at P.A.$=0.82$\D.}
\label{figure:i_img_legend}
\end{figure*}

\begin{figure}
\centering
\includegraphics[width=\columnwidth,angle=0]{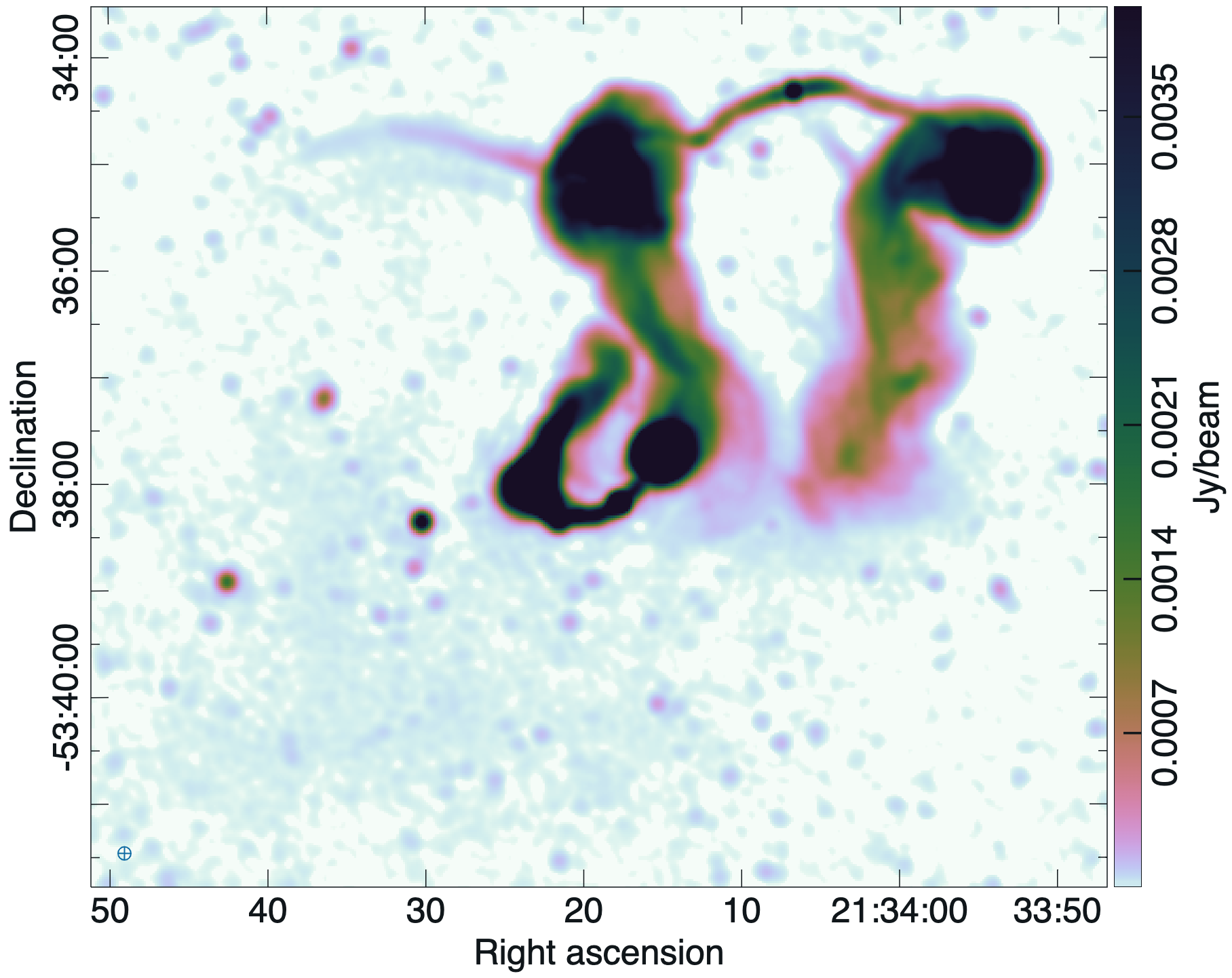}
    \caption{Total intensity image of the Dancing Ghosts obtained with MeerKAT at 
 a nominal frequency of $\nu=1283$~MHz. In the bottom left corner, we show the synthesized beam size of $7.5\times7.1$~arcsec$^2$ at P.A.$=0.82$\D.}
\label{figure:tot_int}
\end{figure}

    Figs.~\ref{fig:DES_ghosts}, \ref{figure:large_field}, \ref{figure:i_img_legend} and \ref{figure:tot_int} show total intensity images of \ghost\ as observed by MeerKAT at a nominal frequency of $\nu =1283$~MHz. \ghost\ is a member of the galaxy cluster Abell~3785, which consists of 47 members and is classified as a compact cluster (type II) by the Bautz-Morgan system, with richness class 0 \citep{BM_system, Abell_clusters}. Host~2 is the \ac{BCG} of this cluster \citep{bcg}. Abell~3785 centre, marked with "X" in Fig.~\ref{figure:large_field}, is positioned at RA$=$21:34:30, DEC$=$--53:37:0 at the redshift of $z=0.0773$ ($D_c \approx 325$~Mpc) and velocity dispersion of $\sigma_v = 897 \pm 312$~km/s \citep{bcg}. There are no previous observations or information on Abell~3785 in X-Ray catalogues.

%%%%%%%%%%%%%%%%%%%%%%%%%%%%%% FIGURE 3 - Large scale cluster %%%%%%%%%%%%%%%%%%%%%%%%%%%%%%
%%%%%%%%%%%%%%%%%%%%%%%%%%%%%% FIGURE 3 %%%%%%%%%%%%%%%%%%%%%%%%%%%%%%

    We consider sources to belong to the Abell~3785 cluster if their position is within a cluster radius, defined by \cite{Abell_clusters} as $R_A = (1.7/z_A) =22$~arcmin (dashed circle in Fig.~\ref{figure:large_field}), and their redshift is between $z=0.055$ to $z=0.095$. We detect 30 sources whose redshifts \citep{abell_6dF,abell_members} and spatial position fit these criteria. These thirty members of the Abell~3785 cluster, including the Dancing Ghosts, are marked as red circles in  Fig.~\ref{figure:large_field}. We also note a deficit of cluster members west of the Dancing Ghosts. 

    %%%%%%%%%%%%%%%%%%%%%%%%%%%%%% FIGURE 4 - legend  %%%%%%%%%%%%%%%%%%%%%%%%%%%%%%

%%% fig5

    The Dancing Ghosts complex spreads over $\approx7.1$~arcmin, corresponding to $\approx624$~kpc at the distance of Abell~3785.  There is a region of diffuse emission to the bottom left (southeast) of the \ghost\ (marked with the blue polygon in Fig.~\ref{figure:large_field}, bottom-right inset). This region is discussed in subsection~\ref{dif_emission}. In that subsection, we also discuss radio object WISEA~J213541.80-534413.2 as a potential cause of the diffuse emission, which is positioned 12~arcmin east from the Dancing Ghosts (Fig.~\ref{figure:large_field} top-left). Finally, two other interesting Abell~3785 members with extended emission  (Fig.~\ref{figure:large_field} top-right) are positioned $17$~arcmin north of the Dancing Ghosts and discussed in more detail in Subsection~\ref{kidney}.

        \subsection{Morphology}
        \label{subsec:ghost}
        \label{subsubsec:morphology}
            \citet{EMU_Norris} established that radio emission from \ghost\ is most likely coming from two \ac{AGN}, northern (Host~1) and southern (Host~2), marked in Fig.~\ref{figure:i_img_legend} as H1C and H2C.

            Both the northern and southern sources have symmetric and narrow inner jets (marked in Fig.~\ref{figure:i_img_legend} as H1J and H2J). After the initial linear flow, jets of the northern host (Host~1) form large lobes (H1L-E and H1L-W). We see this effect in a number of other \ac{AGN}  \citep{RG_review}.
        
            Lobes associated with the northern source extend southwards, forming a low surface brightness structure and wide tails that are marked in Fig.~\ref{figure:i_img_legend} as H1WT-E and H1WT-W. 
            We also note that both lobes and right wide tail (H1L-E, H1L-W, H1WT-W) show multiple low surface brightness filaments (or wisps; see Fig.~\ref{figure:i_img_legend} marked as H1Wisp1 to H1Wisp4) extending in the northeast direction. 
        
            Relatively close (15~arcsec) to the core of the Host~2 (H2C), we see the eastern jet of Host~2 forming a lobe (H2L-E), which bends as it propagates towards Host~1 (H2WT-E). The Host~2 western jet propagates a relatively short distance before forming a circular lobe (H2L-W). The distance to the lobe appears shorter than in the eastern jet. There is some indication of the jet propagation after the lobe, but it is surrounded by the wide tail from the Host~1 (H1WT-E).
    
\begin{figure*}
\centering
\includegraphics[width=.93\textwidth]{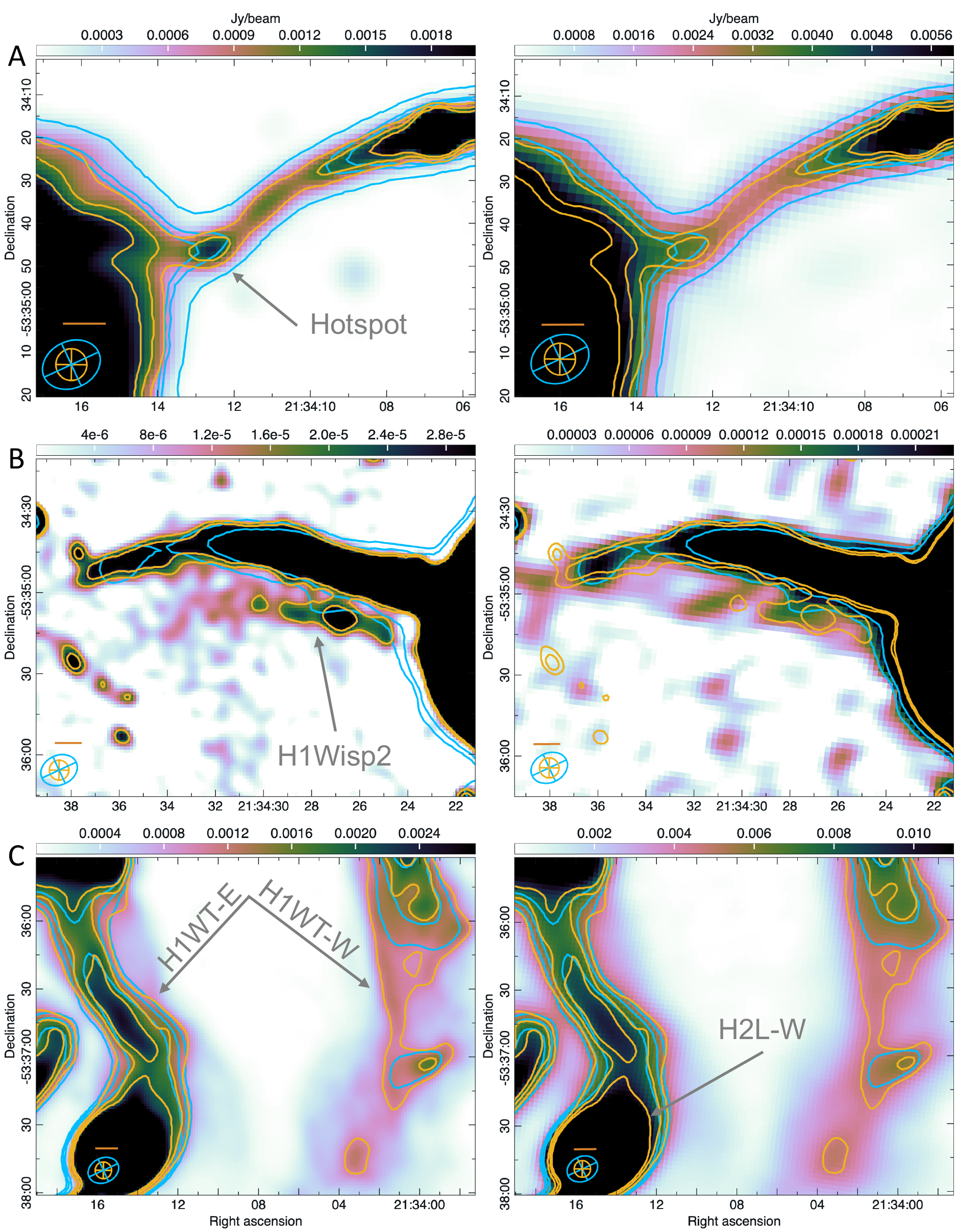}
\caption{Dancing Ghosts subset images corresponding to regions marked as A, B and C on Fig.~\ref{figure:i_img_legend}. Subset images from MeerKAT (left column) and \ac{ASKAP} (right column) are overlaid with orange (MeerKAT) and cyan (ASKAP) contours. In the bottom left corner, we show synthesized beam sizes of $13.95\times10.90$~arcsec$^2$ at P.A.$=-58.15$\D\ as blue ellipses from \ac{ASKAP}, and  $7.5\times7.1$~arcsec$^2$ at P.A.$=0.82$\D\ as orange ellipses from MeerKAT data. The line above the synthesized beam sizes has an angular size of 10~arcsec. Contour levels for each panel are as follows: {\em top panel}: Orange ($0.000016, 0.000025$)~Jy~beam$^{-1}$, Cyan ($0.00015, 0.000275$)~Jy~beam$^{-1}$, {\em middle panel}: Orange ($0.0008$, $0.0012$, $0.0016$)~Jy~beam$^{-1}$, Cyan ($0.0015, 0.003, 0.0035$)~Jy~beam$^{-1}$, {\em bottom panel}: Orange ($0.001$, $0.0015$, $0.002$)~Jy~beam$^{-1}$, Cyan ($0.005,0.006,0.00756.5$)~Jy~beam$^{-1}$.
}
\label{figure:comparison}
\end{figure*}

            In Fig.~\ref{figure:comparison} we compare MeerKAT (left column) with \ac{ASKAP} (right column) images of \ghost. We zoom into three specific regions, marked in Fig.~\ref{figure:i_img_legend} as A, B and C. Notably, MeerKAT observations have increased sensitivity and resolution compared to the \ac{ASKAP} observations. There is an excellent agreement between MeerKAT and ASKAP images, although MeerKAT also detects subtle variations in low surface brightness structures. Orange contours are from the MeerKAT total intensity image, while cyan contours represent total intensity from \ac{ASKAP} image.

            In region A (top panel and  Fig.~\ref{figure:i_img_legend}), MeerKAT detects a hotspot (Fig.~\ref{figure:comparison} top left panel) before the jet reaches the upper left lobe. This is similar to the Hydra~A (3C 218) in the Hydra galaxy cluster \citep[Abell~1060,][]{Hydra_A_Taylor}.
            In region B (middle panel), we see an elongated wisp (H1Wisp1 in Fig.~\ref{figure:i_img_legend}) in both images. In addition, slightly below this wisp, MeerKAT is able to distinguish another low surface brightness feature (wisp, H1Wisp2) with a similar direction and shape. In the third region (C; bottom panel), we see more details and structure of the wide tails (H1WT-E and H1WT-W in Fig.~\ref{figure:i_img_legend}) in the MeerKAT image. H1WT-E is a region where there may be an interaction between the jets originating from different hosts.
 %%%%%%%%%%%%%%%%%%%%%%% FIGURE 6 - TOTAL INTENSITY comparison %%%%%%%%%%%%%%%%%%%%%%%%%%%%%%
%%%%%%%%%%%%%%%%%%%%%%%%%%%%%%%%%%%%%%%%%%%%%%%%%%%%%%%%%%%%%%%%%%%%%%%%%%%%%%%%%%%%%
        \subsection{Spectral Index}
        \label{subsubsec:si}
     
\begin{figure*}
\centering
\includegraphics[width=\textwidth,angle=0]{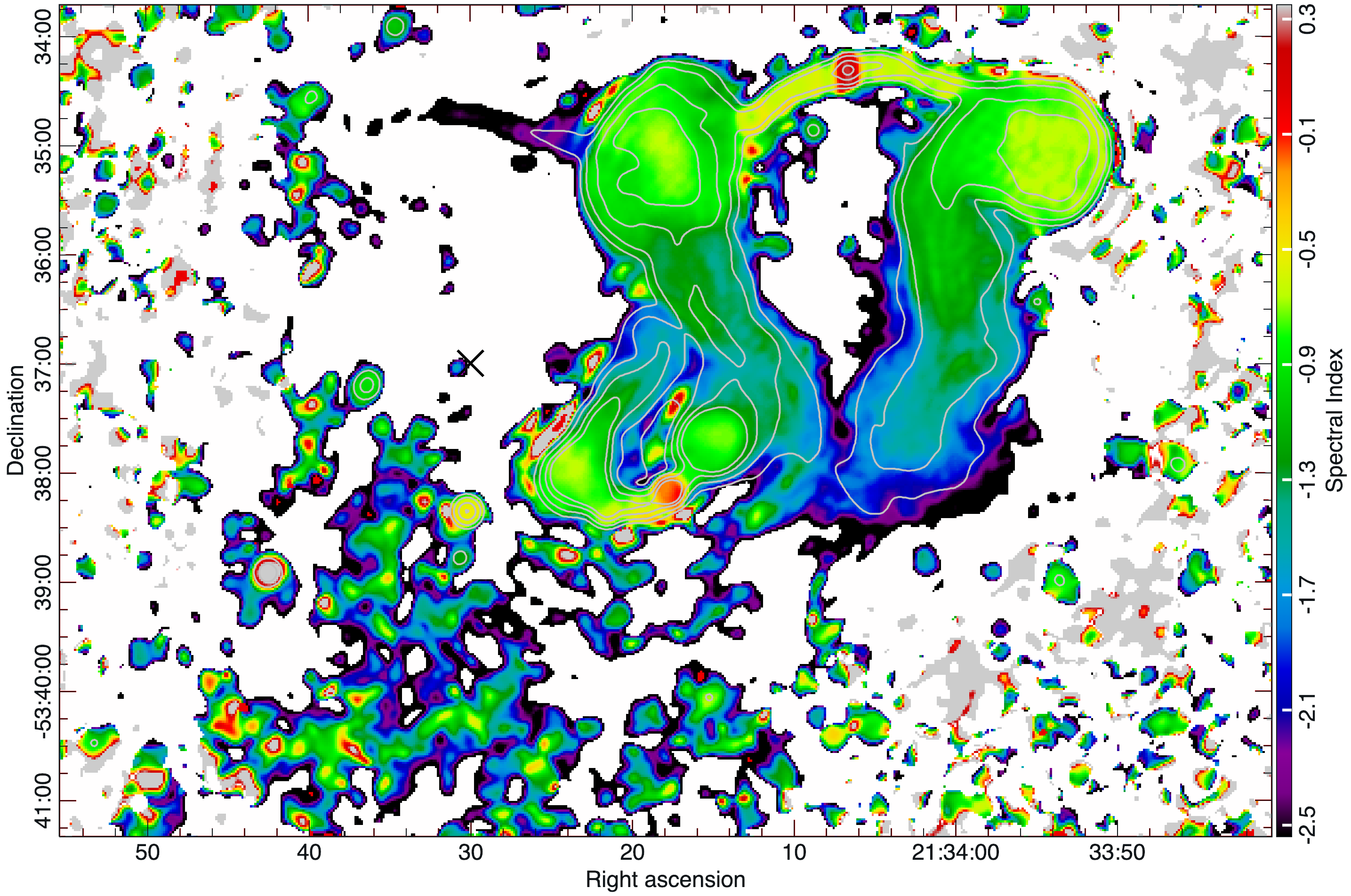}
\caption{Spectral index map created using MeerKAT's $800$~MHz bandwidth images with Brigg's robust weighting of $0$ centred at $\nu=1284$ overlaid with grey contours from total intensity image (0.00015, 0.00055, 0.0015, 0.002, 0.0045 Jy~beam$^{-1}$). The synthesized beam size is ($8.8 \times 7.6$~arcsec$^2$) at P.A.$=-33.5$\D.}
\label{figure:spec_ind}
\end{figure*}

            Fig.~\ref{figure:spec_ind} shows the spectral index image of \ghost\ calculated using MeerKAT images in 14 sequential channels from 886~MHz to 1681~MHz. The details of channelization are given in Table~\ref{tab:channels}. The spectral index value is presented if the reduced chi-squared statistic ( $\chi^2_\nu = \chi^2 / \nu$ ) when fitting the spectral index is no worse than the straight average assuming the default spectral index (0).
    
            We measured flux densities using the method described in \citet[][]{LMC_ORC}. We carefully selected regions that exclude all obvious point sources and measured the total radio flux density, accounting for the local background of a region with {\sc polygon} selection package from the \ac{CARTA} \citep{carta}. We note the selected regions have low surface brightness and/or are sometimes embedded in other complex environments. This may significantly influence the accuracy of our measurements. 
            Similarly, we used the same method and regions to estimate the average spectral index using the image shown in Fig.~\ref{figure:spec_ind}. We estimate that our flux density measurements have an overall scale error of $\approx10$~per~cent of the overall flux density scale error. Our measurements are shown in Table~\ref{tab:1}. 

% %%%%%%%%%%%%%%%%%%%%%%%%%%%%%TABLE 1 - Fluxes %%%%%%%%%%%%%%%%%%%%%%%%%%%%%%%%%%%%%%%%%%
\begin{table*}
\begin{center}
    \caption{Flux densities (peak and integrated) of different regions of \ghost, with an estimated uncertainty of $10$~per~cent. In Column~2 we show the region notation marked in Fig.~\ref{figure:i_img_legend}. The spectral index is calculated independently from Fig.~\ref{figure:spec_ind}.}
    \begin{tabular}{ l l c c c c c} 
    \hline
        \makecell{Region} & \makecell{Region\\ Notation as in Fig.~\ref{figure:i_img_legend}}& \makecell{$S_{\rm p\,MeerKAT}$ \\  (Jy beam$^{-1})$} & \makecell{$S_{\rm I\,MeerKAT}$ \\ (Jy)}  &  $\alpha\pm\Delta\alpha$ \\ 
    \hline
        Host~1: core               & H1C    & $0.0072$ & $0.0077$ & $+0.2 \pm 0.1$ \\
        Host~1: jets               & H1J    & $0.0072$ & $0.0294$ & $-0.7 \pm 0.1$ \\
        Host~1: eastern lobe       & H1L-E  & $0.0141$ & $0.4060$ & $-1.0 \pm 0.2$ \\
        Host~1: western lobe       & H1L-W  & $0.0103$ & $0.3498$ & $-0.9 \pm 0.2$ \\ 
        Host~1: eastern wide tail  & H1WT-E & $0.0027$ & $0.0986$ & $-1.7 \pm 0.3$ \\
        Host~1: western wide tail  & H1WT-W & $0.0024$ & $0.0027$ & $-1.8 \pm 0.4$ \\
        Host~1: wisp 1             & H1Wisp1   & $0.0004$ & $0.0027$ & $-2.5 \pm 0.5$ \\
        Host~1: wisp 3             & H1Wisp3   & $0.0002$ & $0.0005$ & $-2.3 \pm 0.2$ \\
\hline
        Host~2: core               & H2C    & $0.0023$ & $0.0292$ & $-0.2 \pm 0.1$ \\
        Host~2: jets               & H2J    & $0.0062$ & $0.0167$ & $-0.7 \pm 0.2$ \\
        Host~2: eastern lobe       & H2L-E  & $0.0381$ & $0.2928$ & $-0.8 \pm 0.1$ \\
        Host~2: western lobe       & H2L-W  & $0.0312$ & $0.2312$ & $-0.9 \pm 0.1$ \\
        Host~2: eastern wide tail  & H2WT-E   & $0.0149$ & $0.0779$ & $-1.6 \pm 0.3$ \\
    \hline
    \hline
    \end{tabular}
    \label{tab:1}
\end{center}
\end{table*}
        
            The spectral index in the cores of the two Hosts (1 and 2) are $\alpha_{\rm H1C} = +0.2$ and $\alpha_{\rm H2C} = -0.3$. \ac{AGN} cores often have a flat or inverted spectral index as a result of free-free absorption and synchrotron self-absorption originating in the presence of optically thick plasma \citep{inverted_spectrum,inverted_spectrum2}.  
        
            Jets propagating from both hosts display a spectral index of $\alpha= -0.7$, extending all the way to the central region of four lobes. In the lobes, the mean spectral index in the central part is $\alpha \approx -0.8$ and steepens towards the edges to $\alpha= -1.1$. Spectral indices in the tails and wisps are even steeper, ranging from $-1.5$ to under $\alpha \le -2.2$. Except in the AGN cores, the spectral index values calculated using MeerKAT data are slightly steeper than the spectral index from the previous study of the Dancing Ghosts \citep{EMU_Norris} (see Section~\ref{sec:Intro}). However, it is still consistent with bent-tail galaxies in general. 
         
            The New MeerKAT image reveals more of the low surface brightness with enough \ac{SNR} to calculate the spectra in the tails below $-1.5$,  which was noted as a limit in the \ac{ASKAP} observation by \cite{EMU_Norris}. In the regions of jets and lobes, we see a consistency of spectral index between the two observations, within the range of uncertainties. A potential reason for a steeper spectral index in the new image might be the wider bandwidth of the MeerKAT telescope, which is 860~MHz, compared to the \ac{ASKAP} observation with a bandwidth of 288~MHz.
            
            From the four detected wisps shown in Fig.~\ref{figure:i_img_legend}, we estimate the spectral index from wisp~1 and wisp~2 to be $\alpha \le -2.2$. This is similar to what \citet{threads} found in ESO~137--006 as $\alpha \approx -2$.
    
%%%%%%%%%%%%%%%%%% FIGURE 7 - SPECTRAL INDEX %%%%%%%%%%%%%%%%%%%%%%%%%%%%%%
 %%%%%%%%%%%%%%%%%%%%%%%%%%%%%%%%%%%%%%%%%%%%%%%%%%%%%%%%%%%%%%%%%%%%%%%%%%
        \subsection{Polarization and Faraday Rotation Analysis}
        \label{subsubsec:pol}
            We produced the fractional polarization (FPol) using a peak Faraday de-rotated polarized intensity (PPol) map and Stokes I image at nominal frequency $\nu=1283$~MHz. Each pixel in the polarized intensity cube was required to have at least a $5\sigma$ value in one or more planes else the pixel was blanked. We then calculated the fractional polarization as a ratio between the two maps $FPol = PPol / I$.
            
\begin{figure*}
\includegraphics[width=\textwidth]{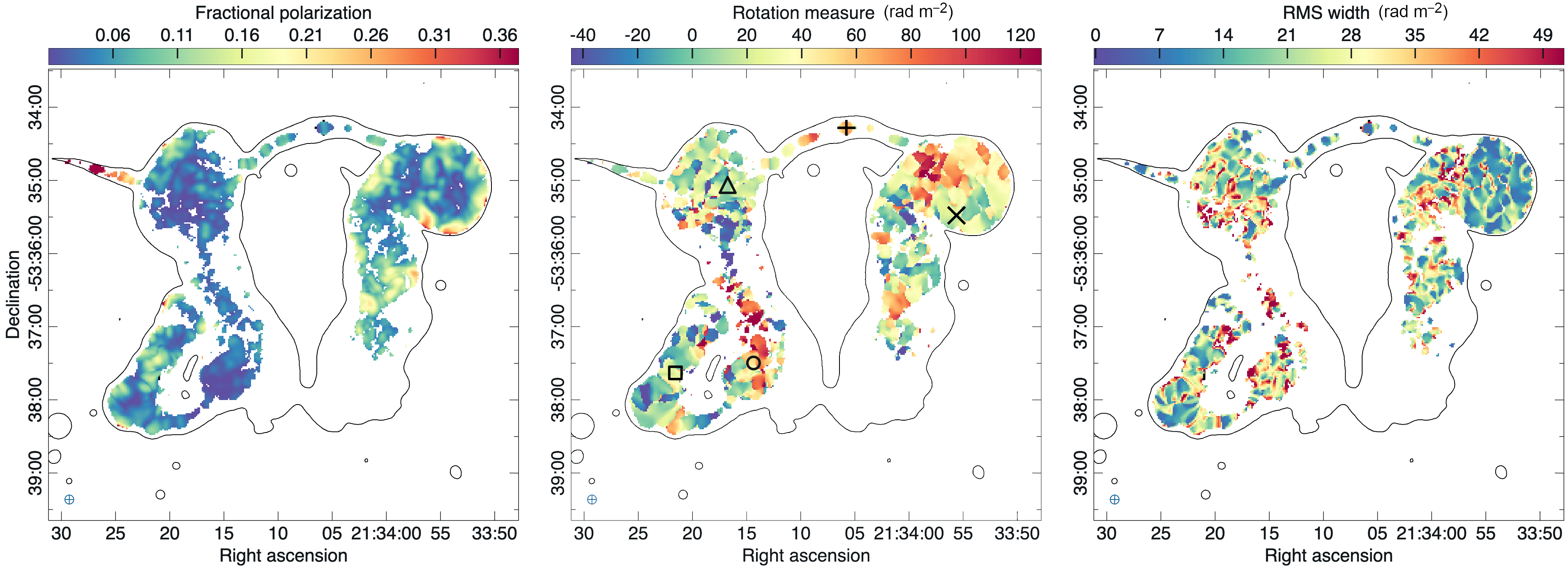}
\caption{\emph{Left}: Fractional polarization map of the Dancing Ghosts at the nominal frequency of 1283~MHz. \emph{Middle}: Peak Rotation Measure map of the Dancing Ghosts created using MeerKAT polarization cubes. Black symbols indicate where plots of the Faraday spectra are given in Fig.~\ref{plots:faraday}. Positive values of RM represent a magnetic field oriented towards the observer. \emph{Right:} RMS width of the feature in the Faraday spectrum in rad~m$^{-2}$ colour-coded by the colour bar at the top. The RMS width of the restoring function used in the CLEAN deconvolution is $6$ rad~m$^{-2}$.
All three images were overlaid with black contour from the total intensity image at $20\sigma$, with $\sigma=5.5$~$\mu$Jy~beam$^{-1}$ to outline the source. The synthesized beam size is shown in the bottom left corner as a blue ellipse ($7.5 \times 7.5$~arcsec$^2$) at P.A.$=-5.71$\D.}
\label{figures:RMFig}
\end{figure*}

            The leftmost panel in Fig.~\ref{figures:RMFig} shows the fractional polarization map of the Dancing Ghosts overlaid with total intensity contours.
            The fractional polarization of all jets in the Dancing Ghosts structure is below $10$~per~cent. We note that regions of H1WT-E, H2L-W and the lower part of H1L-E have low fraction polarization, about $4-5$~per~cent.  Other lobes and wide tails have varying fractional polarization up to $20$~per~cent, which is common in AGN with bent-tail jets \citep{pol_study}. The low integrated fractional polarization appears to be the result of a complex mixture of emitting regions and Faraday screens.
                       
            Interestingly, we detect the highest fractional polarization within a wisp H1Wisp1 on the northeast side and H1Wisp3 reaching up to $50$~per~cent polarized emission. We will further discuss the polarization and spectral index properties of the H1Wisp1 in Section~\ref{subsec:filaments}.
         
%%% NEW FARADAY ROTATION STUFF 

The peak \ac{RM} cube is given in the middle panel of Fig.~\ref{figures:RMFig}. This figure shows a patchy structure covering a wide range of Faraday depths and the locations of the sightlines used to obtain Faraday amplitude spectra.

\begin{figure*}
% \begin{subfigure}{0.48\textwidth}
% \includegraphics[width=2.7in,angle=-90]{figures/polarization/rm/fig9.a_RM_AGN.eps}
%  \label{fig:a}
% \end{subfigure}\hspace*{\fill}
% \begin{subfigure}{0.48\textwidth}
% \includegraphics[width=2.7in,angle=-90]{figures/polarization/rm/fig9.b_RM_North_core.eps}
%  \label{subfigure:RMcore}
% \end{subfigure}
% %\medskip
% \begin{subfigure}{0.48\textwidth}
% \includegraphics[width=2.7in,angle=-90]{figures/polarization/rm/fig9.c_RM_North_ELobe.eps}
% \label{fig:c}
% \end{subfigure}\hspace*{\fill}
% \begin{subfigure}{0.48\textwidth}
% \includegraphics[width=2.7in,angle=-90]{figures/polarization/rm/fig9.d_RM_North_WLobe.eps}
% \label{fig:d}
% \end{subfigure}
% %\medskip
% \begin{subfigure}{0.48\textwidth}
% \includegraphics[width=2.7in,angle=-90]{figures/polarization/rm/fig9.e_RM_South_ELobe.eps}
%  \label{fig:e}
% \end{subfigure}\hspace*{\fill}
% \begin{subfigure}{0.48\textwidth}
% \includegraphics[width=2.7in,angle=-90]{figures/polarization/rm/fig9.f_RM_South_WLobe.eps}
% \label{fig:f}
% \end{subfigure}
\includegraphics[width=\textwidth]{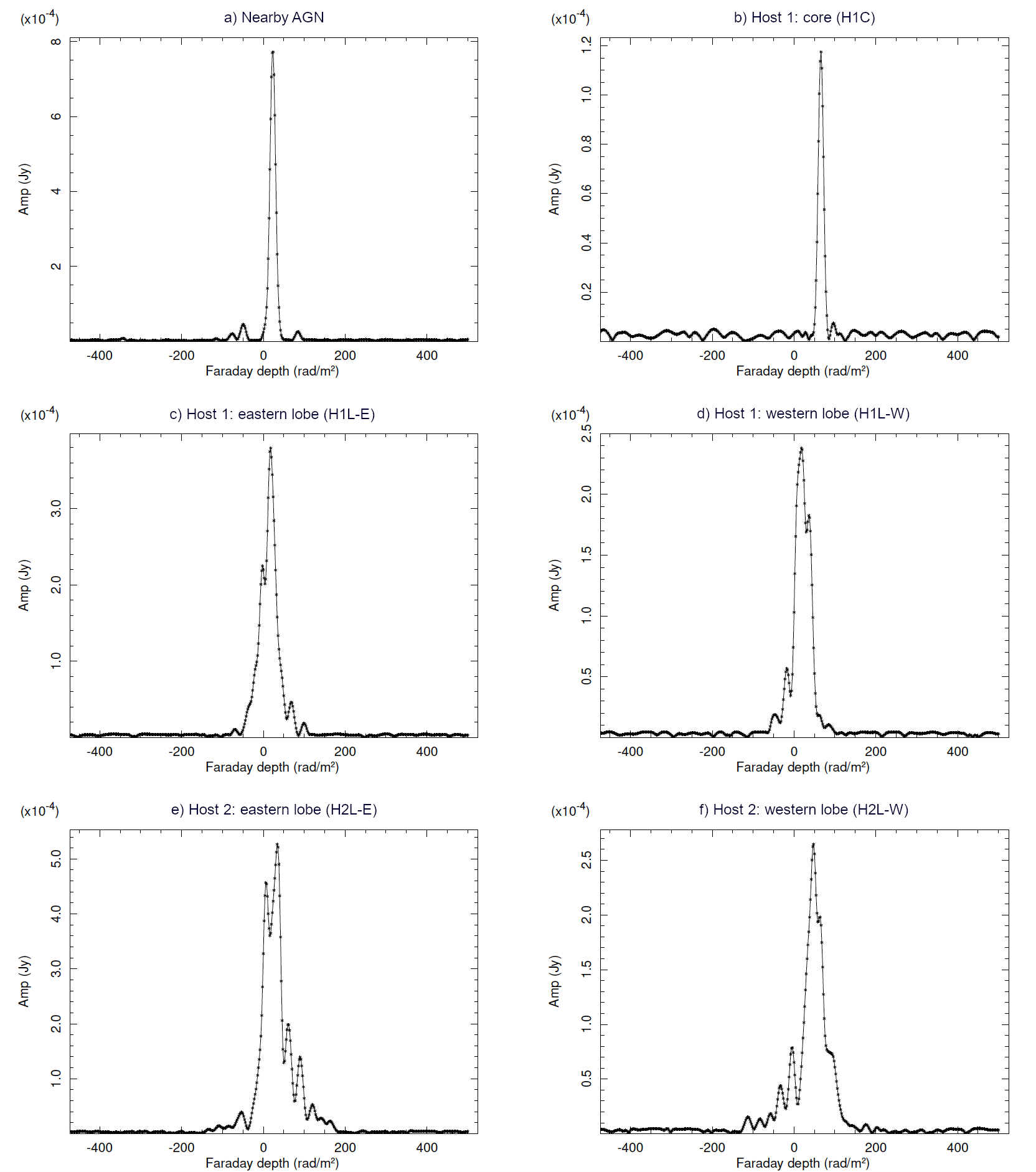}
\caption{Faraday amplitude spectra with the exact location of sightlines indicated with black symbols in the middle panel of Fig.~\ref{figures:RMFig}.
a) Nearby AGN, R.A.$=21$\hour$35$\minute$47.98$\second, Dec$=-53$\D $28$\arcmin $44.0$\arcsec,
b) polarized knot in inner jet, near the core of northern source (H1C), ``+'' in Fig.~\ref{figures:RMFig},
c) eastern lobe of northern source (H1L-E), triangle in Fig.~\ref{figures:RMFig},
d) western lobe of northern source (H1L-W), ``X'' in Fig.~\ref{figures:RMFig},
e) eastern lobe of southern source (H2L-E), square in Fig.~\ref{figures:RMFig},
f) western lobe of southern source (H2L-W), circle in Fig.~\ref{figures:RMFig}. Sightlines through the lobes of the Dancing Ghosts (panels: 9c - 9f) show components with thick Faraday structures spreading over 100 rad/m$^{2}$.}
\label{plots:faraday}
\end{figure*}

Fig.~\ref{plots:faraday} shows the Faraday amplitude spectra at a number of sightlines. These sightlines through the inner sources are indicated by the black symbols in the middle panel of Fig.~\ref{figures:RMFig}. Spectra on a nearby AGN (Fig.~\ref{plots:faraday}a) and a polarized knot in the inner jet of the northern source (Fig.~\ref{plots:faraday}b) show simple, single, unresolved Faraday components. Sightlines through the lobes (Figs.~\ref{plots:faraday}c -- \ref{plots:faraday}f) show multiple components spread over up to 100 or more rad\,m$^{-2}$. These Faraday thick structures will certainly contribute to the depolarization seen in the extended regions of the central sources in Fig.~\ref{figures:RMFig} 

\subsubsection{Faraday Width}

Many of the sightlines whose Faraday spectra are shown in Fig.~\ref{plots:faraday} show an extended response in Faraday depth. The \ac{RMS} width of the Faraday response in pixels with significant polarized emission is given in Fig.~\ref{figures:RMFig}. Note, the \ac{RMS} width of the restoring function used in the CLEAN is about 6 rad\,m$^{-2}$. 
Requirements for pixels in the rightmost panel in Fig.~\ref{figures:RMFig} to have a non-blank value must meet certain criteria. The pixel must have at least one plane with a value greater than $25\, \times \,$\ac{RMS} value of the plane, and it must have more than $20$~per~cent of the peak value in the spectrum in that pixel. Additionally, the maximum value in the spectrum for each pixel must be above $30\,\mu$Jy and the integral values over non-blanked pixels must exceed $100\,\mu$Jy. The presented width in Faraday depth will substantially depolarize the emission.

\subsubsection{Implications of the Dense Faraday Screen}
 
The widespread low fractional polarization seen in Fig.~\ref{figures:RMFig} shows an unusually large depolarization of the extended emission. The source of this depolarization is likely the dense Faraday screen in front of this emission. The chaotic nature of the thick Faraday screen seen in Figs.~\ref{figures:RMFig} and \ref{plots:faraday} suggest a strong
interaction between the radio sources and the cluster medium with small scale variations in the foreground Faraday rotating medium. Since the Faraday rotation is proportional to the product of the magnetic field and the thermal electron density, this fine-scale structure could be from variations in the magnetic field, electron density, or both. The sightline to a polarized knot in the inner jet of the northern source (Fig~\ref{plots:faraday}b) shows a relatively simple Faraday spectrum. This might be due to a foreground Faraday screen with a characteristic scale, smaller than the beamwidth but larger than the jet dimensions. If this is the case, the emission from a bright, unresolved jet knot then all sees the same Faraday depth, so the peak in the Faraday spectrum appears unresolved. Reversals in the sign of the Faraday depth indicate reversals in the magnetic field direction.

The complex Faraday screens are similar to that seen in other Abell clusters such as Abell~3667 \citep{abell_discussion}. This suggests that the radio emission from the \ghost\ pair of galaxies is largely shaped by the interaction of the jets with the \ac{ICM} in the host cluster, Abell~3785. However, additional X-ray observations and analysis of Abell~3785 are needed to confirm the jet-\ac{ICM} interaction.

We also obtained the expected Galactic foreground rotation measure obtained using the CIRADA cutout server\footnote{\url{http://cutouts.cirada.ca/rmcutout/}}. At the position of the Dancing Ghosts, the Galactic foreground RM is $15 \pm 6$~rad~m$^{-2}$.

\section{Discussion}
\label{sec:discussion}

    \subsection{Signs of interaction in the Dancing Ghosts?}
    \label{subsec:interactions}

        The projected distance between Host~1 and Host~2 is about $d\approx 369$\,kpc (4.2\,arcmin) in the plane of the sky. Because the only available redshifts are photometric with a typical uncertainty of $\delta z \approx 0.01=3000$\,km/s,  the peculiar velocity difference of $\approx580$\,km\,s$^{-1}$ inferred from redshifts is not significant. We, therefore, have no information on the radial separation of the two hosts. 
        
\begin{figure}
\centering
\includegraphics[width=\columnwidth]{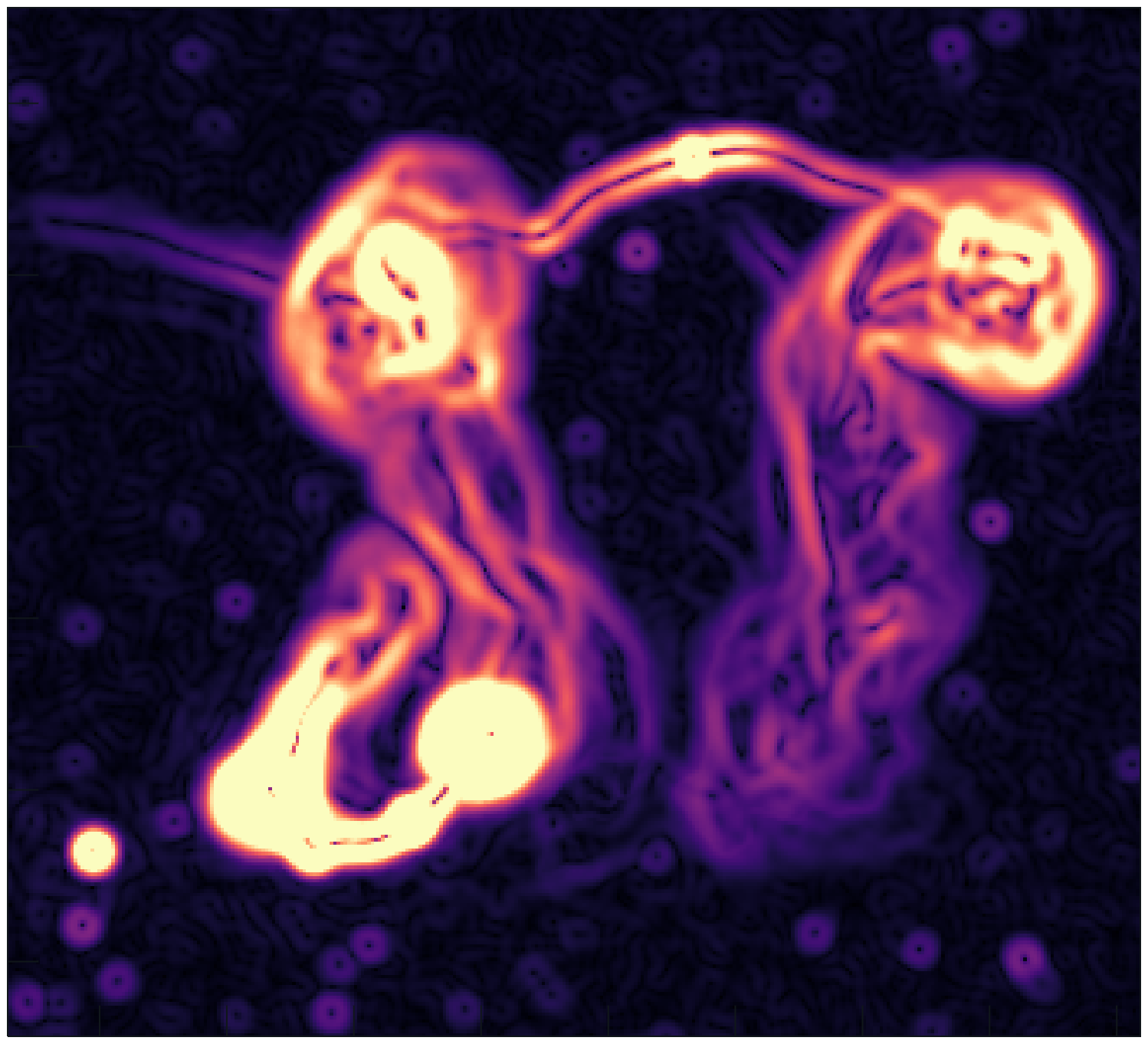}
\caption{Image generated by a Sobel edge-detection filter of the Dancing Ghosts, to emphasize the filamentary structures in radio galaxy complex. This was generated using the {\sc niner} task in {\sc aips} with synthesized beam size $7.5 \times 7.1$~arcsec$^2$.}
\label{fig:sobel}
\end{figure}

        To better understand the spatial structure of the Dancing Ghosts and their possible interactions, we further analysed spectral index (Fig.~\ref{figure:spec_ind}) and fractional polarization
        maps (Fig.~\ref{figures:RMFig}) and used the Sobel matrix filter on total intensity image (Fig.~\ref{fig:sobel}). We looked for signs of discontinuities, in the form of sharp changes in polarization or spectral index, and filamentary structure within the region of interest (H2L-W and H1WT-E).
    
        We note a slight increase in the polarization intensity and change of magnetic field orientation in the H1WT-E region (Figs.~\ref{figures:RMFig}), but no obvious discontinuities that would indicate possible interaction of the jets. Spectral index profiles along the H1WT-E region are relatively constant at about $\alpha \approx-1.7 \pm 0.1$, and in the transverse cuts, they go from $\alpha \approx -1.7$ at the ridge line to $\alpha \approx -2.3$ at the edges.

         Additionally, in Fig.~\ref{fig:sobel} we note a number of filamentary structures, inside the Dancing Ghosts galaxy pair. Filaments in the above-mentioned region, H1WT-E, do not show signs of intertwining. We also note that Host~1 lobes and wide tails would look symmetric if we subtract the emission from Host~2.    
       
%%%%%%%%%%%%%%%%%%%%%%%%%%%%%% FIGURE 10 - Sobel %%%%%%%%%%%%%%%%%%%%%%%%%%%%%%
%%%%%%%%%%%%%%%%%%%%%%%%%%%%%%%%%%%%%%%%%%%%%%%%%%%%%%%%%%%%%%%%%%%%%%%%%

        After analyzing the region where the interaction is most likely to occur, we do not have strong evidence for the interaction between the galaxy pair. The absence of discontinuities in polarization, spectral index, and total intensity images, as well as inconclusive results from differences in peculiar velocities, suggests that the two galaxies are physically separate. We probably see the \ghost\ complex as a projection effect along the line of sight. Nevertheless, further light may be shed on the level of interaction between the pair by MHD simulations and modelling.

    \subsection{The nature of diffuse emission near the Dancing Ghosts}
    \label{dif_emission}
        In the following subsection, we discuss the diffuse emission marked with a blue polygon in Fig.~\ref{figure:large_field}. It is $\approx$7~arcmin across ($D_A\approx$615~kpc at the distance of Abell~3785) and shows an emission that is stronger than the calibration artefacts and over three times above the local \ac{RMS} ($\sigma \approx 5.5$~$\mu$Jy). 

\begin{figure}
\centering
\includegraphics[width=\columnwidth]{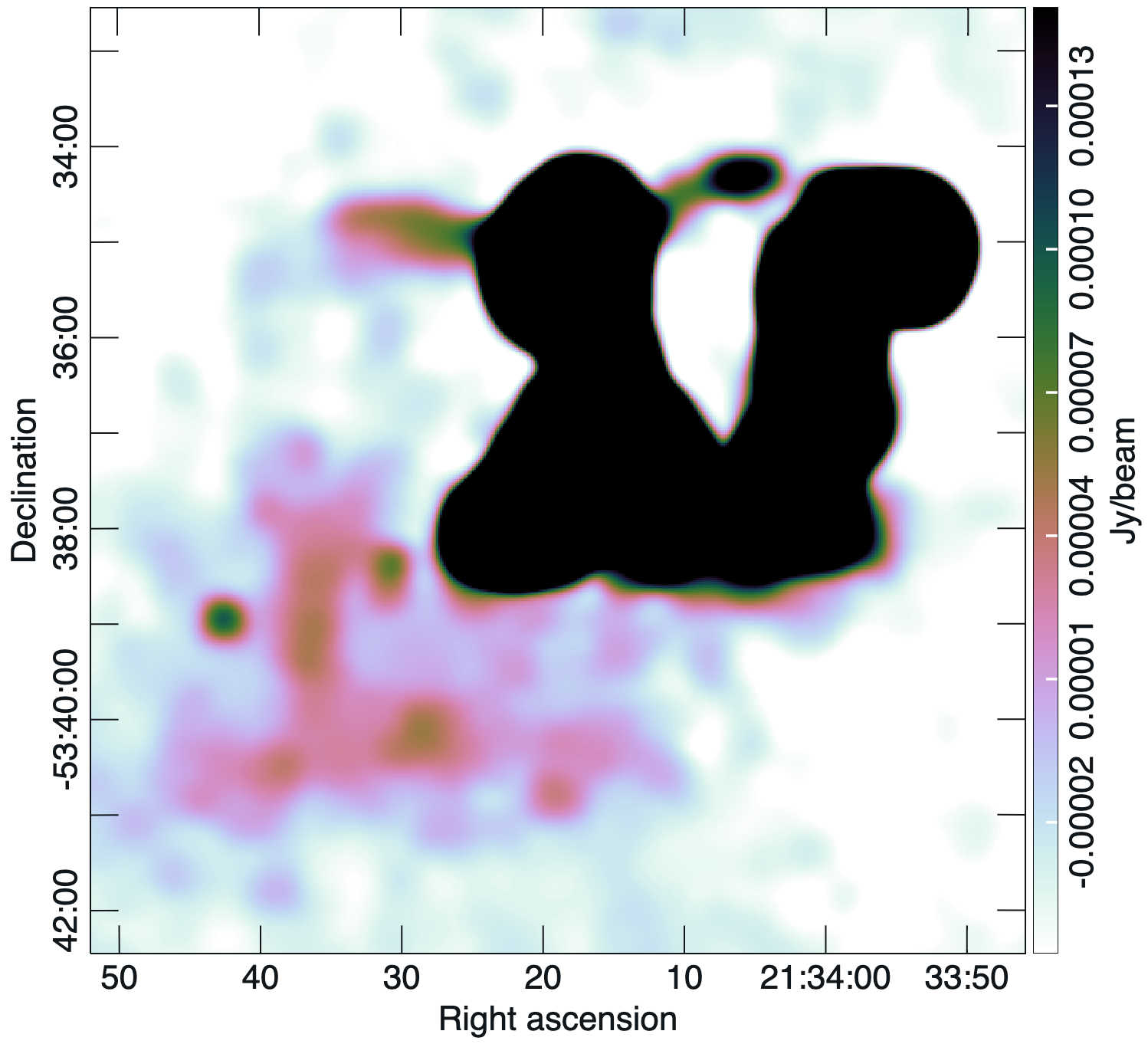}
\caption{Resulting image of the Dancing Ghosts after using the multiresolution filtering method to subtract point sources and enhance diffuse emission from the total intensity image. The synthesized beam size is convolved to $18 \times 18$~arcsec$^2$.}
\label{fig:diffuse}
\end{figure}

        We also used the multiresolution filtering method of \citet{rudnick_mrfilter} to enhance the diffuse emission around the Dancing Ghosts. Fig.~\ref{fig:diffuse} shows the resulting image revealing the low surface brightness emission region southeast of \ghost\ with a distinctive structure. 
      %%%%%%%%%%%%%%%%%%%%%%%%%%%%%% FIGURE 11 - diffuse emission %%%%%%%%%%%%%%%%%%%%%%%%%%%%%%
%%%%%%%%%%%%%%%%%%%%%%%%%%%%%%%%%%%%%%%%%%%%%%%%%%%%%%%%%%%%%%%%%%%%%%%%%

        The origin of the extended and low surface brightness emission is unclear. Here, we present a few possible origins of this diffuse emission. 

        Since the diffuse emission is unpolarized and near the centre of the Abell~3785 cluster and its \ac{BCG} (Host~2), it is possible that the emission represents the cluster's radio halo. The slight offset between the centre (marked X in Fig.~\ref{figure:large_field}) of the cluster and its radio halo emission indicates that it could have suffered a recent merger event, and has not had enough time to relax \citep{mergers_phys, merger_evo}. Future studies of the cluster using X-ray data could indicate if the emission is indeed coming from the halo. 

\begin{figure}
    \centering
        \includegraphics[width=2.75in, angle=-90]{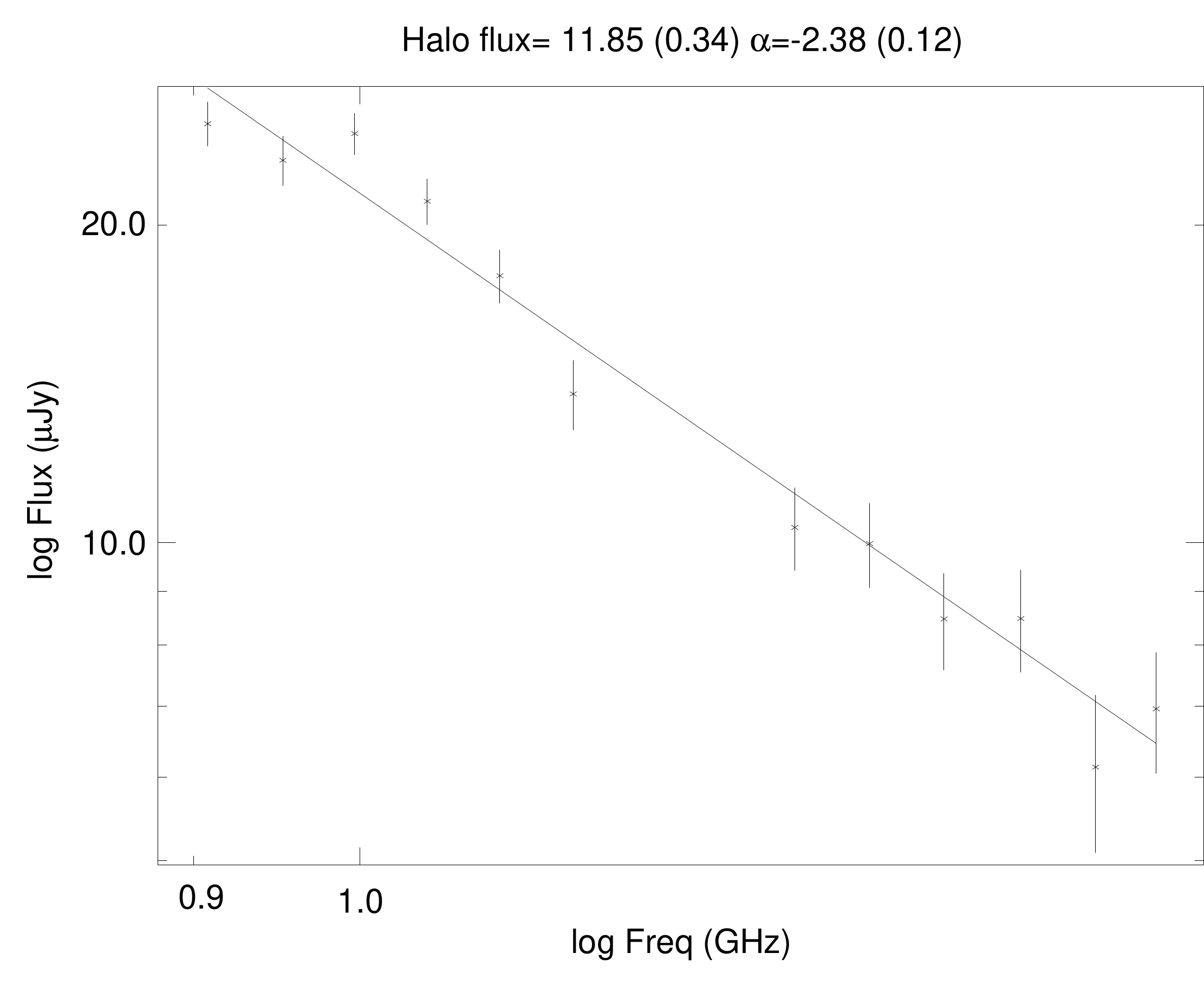}
        \caption{Spectrum of the diffuse emission southeast of the Dancing Ghosts, blue polygon region in Fig.~\ref{figure:large_field}, calculated between 886 and 1681~MHz in 14 sequential channels described in Table~\ref{tab:channels}.}
    \label{fig:si_diffuse}
    \end{figure}

        Another possibility is that the emission is a remnant radio galaxy that is being compressed and distorted by either a merger shock or interaction with either Host~2 or \ac{ICM} winds. If this were the case, the strength of the magnetic field and the momentum of the relativistic electrons in this plasma would increase, resulting in synchrotron emission \citep{diffuse_origin}. 
        To calculate the spectral index of the diffuse emission we used the images with Brigg's "optimal" Robust weighting of $0$ ({\sc AIPS/Obit}) which gives better overall sensitivity, as well as better surface brightness sensitivity at a cost of reduced resolution. In Fig.~\ref{fig:si_diffuse} we show that the integrated spectral index in the diffuse region is very steep $\alpha \approx -2.4$, favouring the relic origin of the emission.

    %%%%%%%%%%%%%%%%%%%%%%%%%%%%%% FIGURE 12 - diffuse emission %%%%%%%%%%%%%%%%%%%%%%%%%%%%%%
    %%%%%%%%%%%%%%%%%%%%%%%%%%%%%%%%%%%%%%%%%%%%%%%%%%%%%%%%%%%%%%%%%%%%%%%%%
    
        For a galaxy to be classified as a remnant or fossil galaxy it needs to have its lobes detached from the AGN core, with no activity in the nucleus or jets \citep{fossils2,fossils}.
        We also note that the region of diffuse emission might be connected to an ancient lobe originating from either of the two \ghost\ hosts.

        Finally, we suggest that this extended emission might originate from a neighbouring galaxy in the Abell~3785 cluster. Just $\approx$15\,arcmin to the south-east, there is another Abell~3785 member WISEA~J213541.80-534413.2 at $z=0.06254$ ($D_c\approx264$~Mpc) that shows an \ac{AGN} jet pointing towards this extended emission and directly via other Abell~3785 member J213522.85--534200.3 at the redshift $z=0.08114$ ($D_c\approx341$~Mpc) (see Fig.~\ref{figure:large_field} top-left inset). The prominent jet of WISEA~J213541.80-534413.2 shows a flat spectral index at its core ($\alpha = 0.15$) and steepens with distance as the jet propagates to $\alpha=-2$. 
        While possible, it is unlikely that the diffuse emission originates from WISEA~J213541.80-534413.2.

        The fact that Host~2 is the \ac{BCG} of the Abell~3785 cluster implies that the emission is most likely affected by shocks, which raises the possibility the origin of the diffuse emission is either from a radio halo or a fossil galaxy. It is important to note that we could not find a cluster's radio halo with a similar shape in the literature.

    \subsection{The nature of wisps}
    \label{subsec:filaments}
              
        Here we discuss the thin, faint features originating from the lobes (marked in Fig.~\ref{figure:i_img_legend} as H1Wisp1-H1Wisp4). These thin synchrotron filaments are emerging new phenomena in radio images with sufficient resolution and sensitivity. They have been detected in recent studies of various galaxy clusters and \ac{AGN} \citep[e.g. Abell 3376, IC 4296, ESO 137-006 by][ respectively]{abell_discussion,condon_ribbons,threads}. Furthermore, \citet{abell_194} thoroughly examine the thin synchrotron filaments and discuss the most likely scenarios of their origins. \cite{Y_Zadeh} compare well-studied Galactic and recently discovered extragalactic synchrotron filaments. They argue that both phenomena originate from the same physical processes but on drastically different scales.  

\begin{figure}
\centering
\includegraphics[width=\columnwidth]{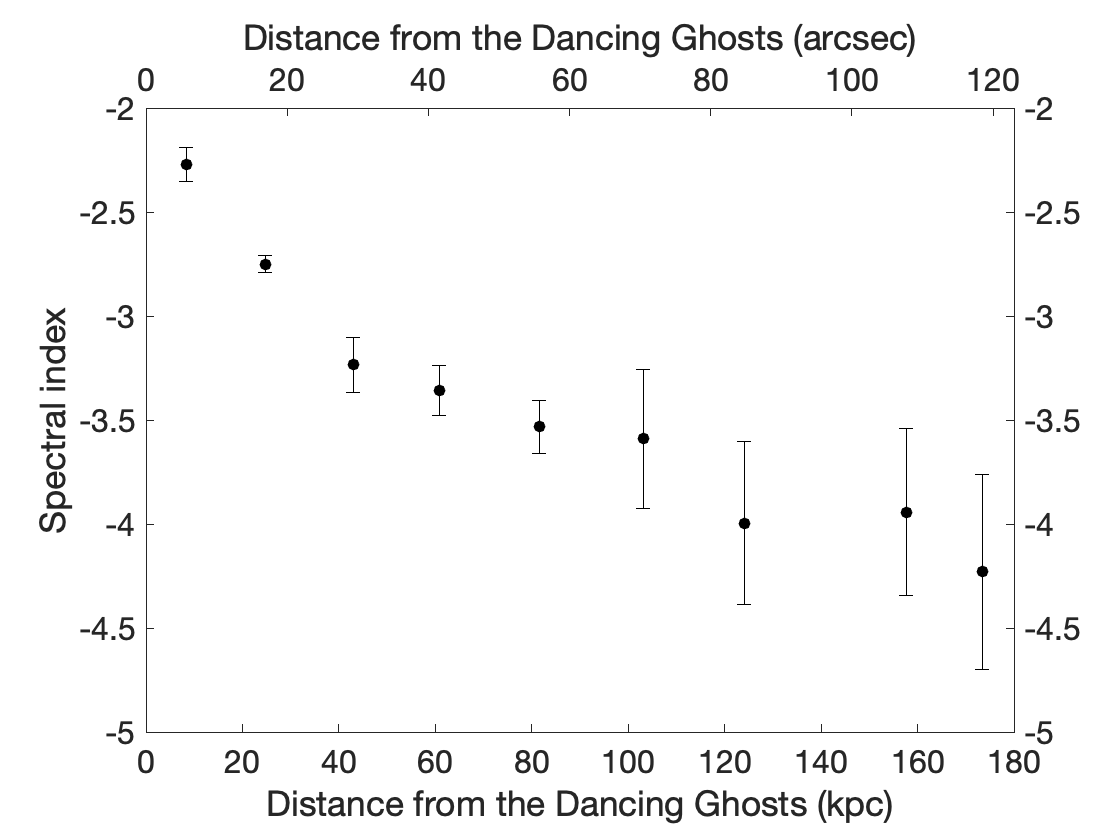}
\caption{Spectral index of the wisp (H1Wisp1) as the function of the distance from the lobe of the Dancing Ghosts. Calculated between 886 and 1681~MHz in 14 sequential channels described in Table~\ref{tab:channels}.}
\label{fig:si_filament}
\end{figure}

        The Meerkat Observations of the Dancing Ghosts and filamentary structure suggest that H1Wisp1 has a broad envelope extending across H1Wisp2. 
        Fig.~\ref{fig:si_filament} shows the spectral index of the H1Wisp1 as a function of the distance from the lobe of the Dancing Ghost. 
        
\begin{figure}
\centering
\includegraphics[width=\columnwidth]{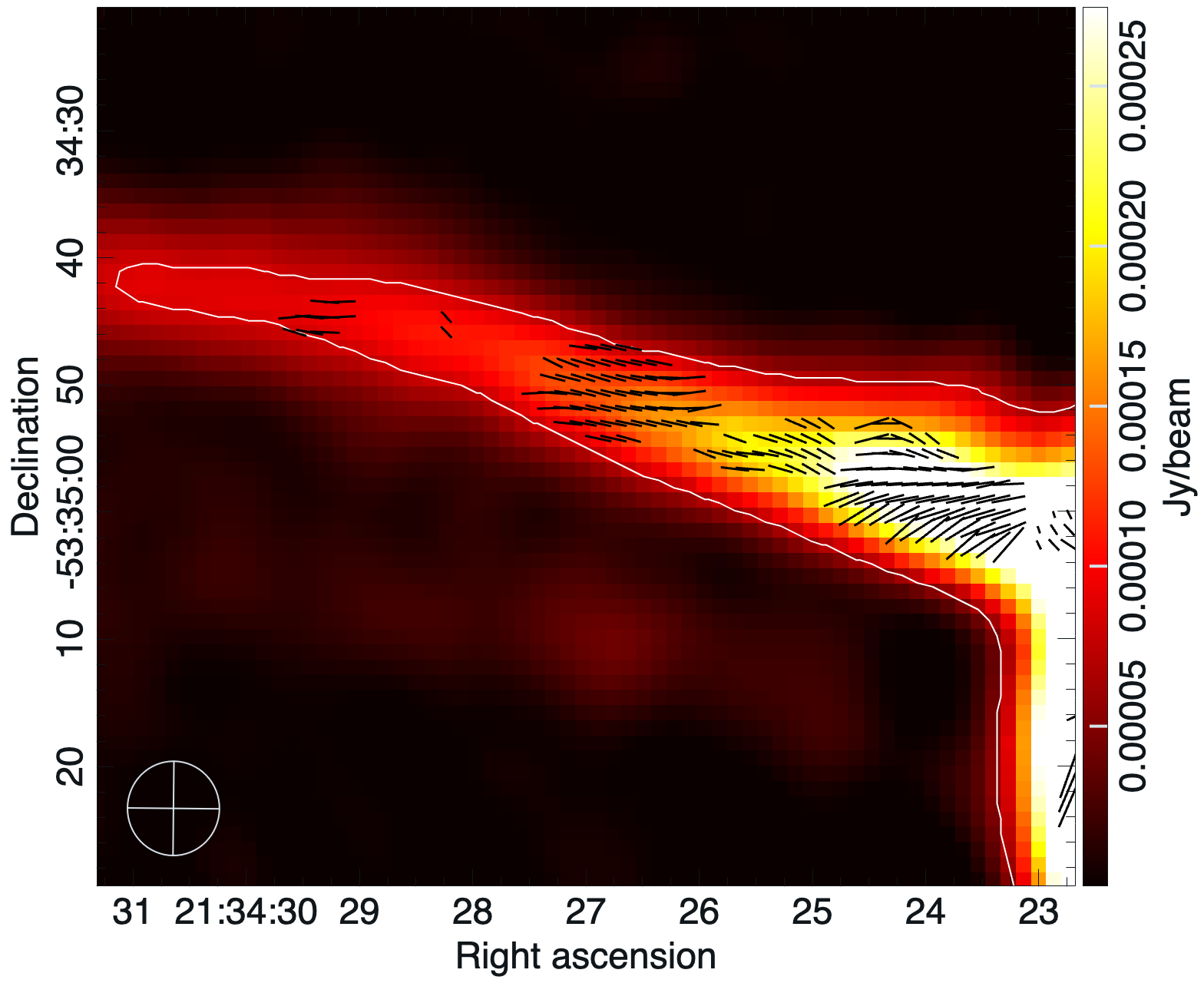}
\caption{Total intensity image of the wisp (H1Wisp1) overlaid with magnetic field vectors after RM correction and the grey contours at $78~\mu$~Jy. The synthesized beam size is shown in the bottom left corner as a white ellipse $(7.5 \times 7.1)$~arcsec$^2$ at P.A.$=0.82$\D.}
\label{fig:pol_filament}
\end{figure}

        Fig.~\ref{fig:pol_filament} shows the magnetic field vectors derived from polarization data and de-rotated for the Faraday rotation effect. Magnetic field vectors are relatively ordered and oriented along the filamentary structure. It should be noted that due to the unresolved Faraday spectra in the region of filaments and our derivation is unambiguous. These characteristics are very similar to what \cite{abell_194} detected in filaments of 3C40B.
        
        The steepening of spectral index could come from two causes, which cannot be easily disentangled:  1) a curved electron spectrum seen in progressively weaker magnetic fields, thus showing steeper spectral indices as the fields get weaker and the radiation samples the higher energy part of the spectrum, and 2) radiative ageing if there is a flow of material from the radio galaxy.

        Systematic changes in brightness and spectrum as a function of distance from the bright radio galaxy structure suggest that the wisp (H1Wisp1) is part of the Dancing Ghosts and not a superposition with a background source.

        Magnetised filaments composed of smaller fibres are created in turbulent, highly pressurised and magnetic plasmas, and could also be further stretched by the jet encounter with the dense cloud moving within \ac{ICM}. However, this scenario does not explain the spectral structure and the flux density change as a function of the distance from the lobe of the Dancing Ghosts.
        
 %%%%%%%%%%%%%%%%%%%%%%%%%%%%%% FIGURE 13 - Wisp SI si%%%%%%%%%%%%%%%%%%%%%%%%%%%%%%
%%%%%%%%%%%%%%%%%%%%%%%%%%%%%%%%%%%%%%%%%%%%%%%%%%%%%%%%%%%%%%%%%%%%%%%%%     

%%%%%%%%%%%%%%%%%%%%%%%%%%%%%% FIGURE 14 -Wisp -polarization %%%%%%%%%%%%%%%%%%%%%%%%%%%%%%
%%%%%%%%%%%%%%%%%%%%%%%%%%%%%%%%%%%%%%%%%%%%%%%%%%%%%%%%%%%%%%%%%%%%%%%%%

        Other explanations include the interaction between the jets and the dense \ac{ICM} cloud, either stationary or moving and cosmic ray acceleration of the diffuse emission originating from the \ghost. 
        
        In the scenario where the jet hits a very dense cloud, we expect deflection -- bending of the jets. The Dancing ghost pair abruptly changes the flow direction near the most prominent wisps, H1Wisp1 and H1Wisp3.
        We note that H1Wisp3 is possibly reconnecting with the core of Host~1, which might be the source of cosmic rays for re-acceleration of electrons within a low-diffuse emission, but for the remaining wisps there has to be interaction with the \ac{ICM}. 
        
        The strong evidence for the existence of intracluster galactic wind and the turbulent \ac{ICM} is that all the wisps originating in the Dancing Ghosts point from the lobes towards the northeast regardless of the jet flow in the local area.

    \subsection{North-west sources of Abell~3785}
    \label{kidney}

  %%%%%%%%%%%%%%%%%%%%%%%%%%%%%% FIGURE 15 - Northern Source %%%%%%%%%%%%%%%%%%%%%%%%%%%%%%
\begin{figure}
\centering
\includegraphics[width=\columnwidth]{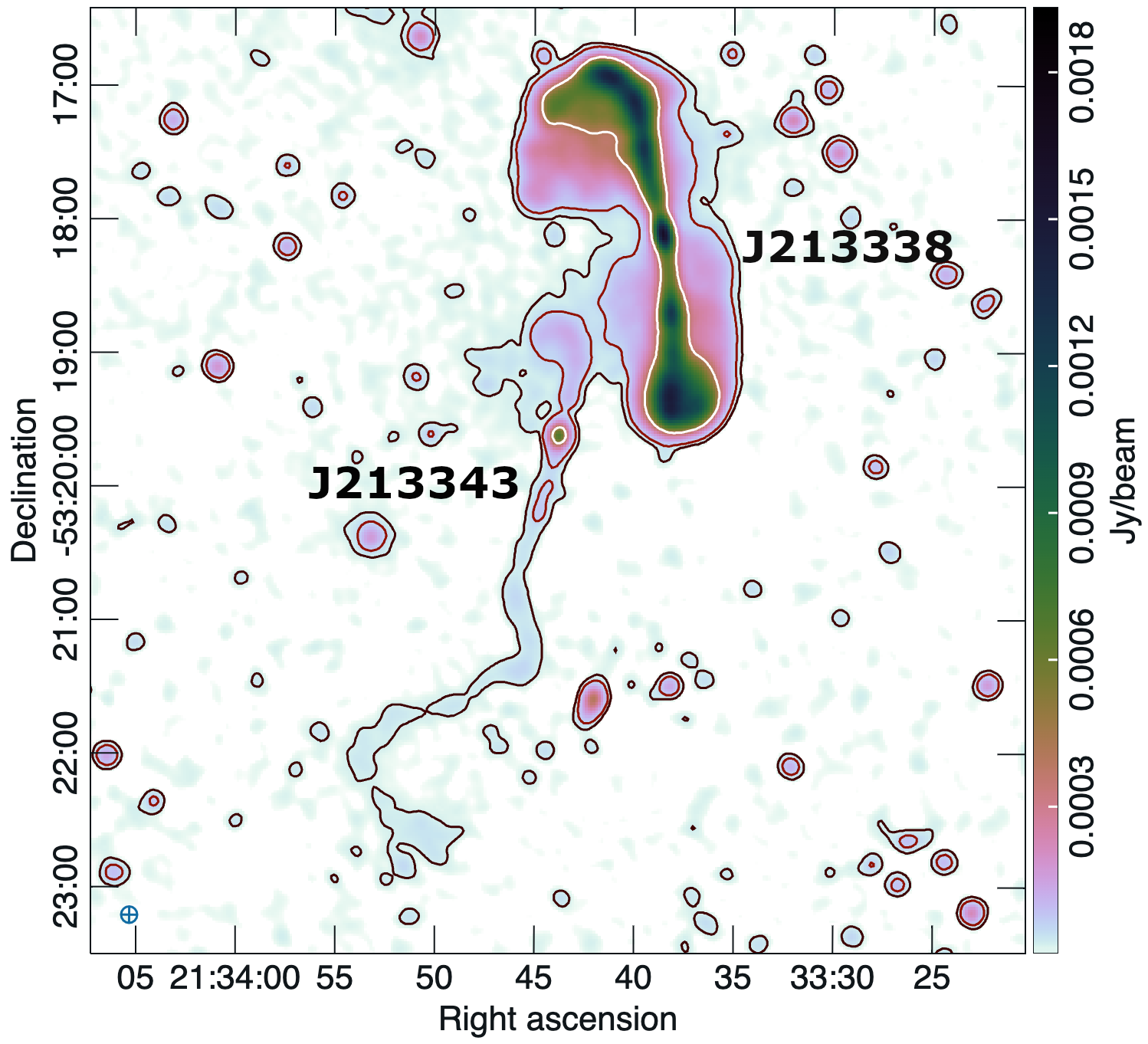}
\caption{
      Image of the two extended radio sources within Abell~3785 and NW of the Dancing Ghosts. Contour levels are at $[3,10,85]\times \sigma$, with $\sigma = 4.78$~$\mu$Jy~beam$^{-1}$, presented from dark to light colour. The synthesized beam size is shown in the bottom left corner as a blue ellipse ($7.5 \times 7.1$~arcsec$^2$) at P.A.$=0.82$\D.
      }
\label{fig:kidney}
\end{figure}
%%%%%%%%%%%%%%%%%%%%%%%%%%%%%%%%%%%%%%%%%%%%%%%%%%%%%%%%%%%%%%%%%%%%%%%%%
    
    The two brightest radio sources show clear two-sided \ac{AGN} jets, which are pointing towards \ghost\ (Fig.~\ref{fig:kidney}). These two are WISEA~J213338.54--531806.2  ({\sc J213338}; at $z = 0.0751$ or $D_c\approx316$~Mpc) and WISEA~J213343.77--531937.1 ({\sc J213343}; at $z=0.0826$ or $D_c\approx347$~Mpc). 
    
    The two above-mentioned extended objects have steep radio spectra. {\sc J213338} has a spectral index at the core $\alpha=-1$ and relatively constant along the jets $\alpha=-1.2$ but steepening to $\alpha \le -2$ at the edges of the jets, while {\sc J213343} on the other hand is relatively steep at the core $\alpha = -0.76$ and afterwards intermittently flattens and steepens in a twisted and collimated jet.

\section{Conclusion}
\label{sec:conclusion}

We use high-resolution and high-sensitivity MeerKAT observations to study the galaxy cluster Abell~3785 and its peculiar galaxy pair -- \ghost, the Dancing Ghosts, previously detected in the \ac{ASKAP} \ac{EMU} pilot survey. 

\begin{itemize}
    \item MeerKAT observations reveal more detailed low surface brightness structures, filamentary structures and a hotspot near the northern left lobe.
    \item Of particular interest are a number of wisps emanating from the lobes of the radio galaxies. We do not currently have a satisfactory explanation of these.
    \item Fractional polarization and Faraday analysis show a very complex magnetic field with a thick and dense Faraday screen resulting in a large depolarization of the extended emission within the Dancing Ghosts, which is also affected by \ac{ICM}. 
    \item We see no sign of interaction between the two radio galaxies in the Dancing Ghosts, and, although both are associated with the cluster, we suggest they may merely be superimposed  along the line of sight.
    \item Diffuse emission near the Dancing Ghosts may represent a radio halo of the Abell~3785 cluster or a fossil radio galaxy.
    \item We detect 30 radio continuum sources within Abell~3785 including the Dancing Ghosts.
\end{itemize}

\section*{Acknowledgements}

We thank Lawrence Rudnick for providing useful discussions and insights which greatly contributed to this paper.

The MeerKAT telescope is operated by the South African Radio Astronomy Observatory, which is a facility of the National Research Foundation, an agency of the South Africa Department of Science and Innovation.

The National Radio Astronomy Observatory is a facility of the US National Science Foundation, operated under a cooperative agreement by Associated Universities, Inc.

The Australian SKA Pathfinder is part of the Australia Telescope National Facility which is managed by the \ac{CSIRO}. The operation of \ac{ASKAP} is funded by the Australian Government with support from the National Collaborative Research Infrastructure Strategy (NCRIS). \ac{ASKAP} uses the resources of the Pawsey Supercomputing Centre. The establishment of \ac{ASKAP}, the Murchison Radio-astronomy Observatory and the Pawsey Supercomputing Centre are initiatives of the Australian Government, with support from the Government of Western Australia and the Science and Industry Endowment Fund.
We acknowledge the Wajarri Yamatji people as the traditional owners of the Observatory site.

This research has made use of the NASA/IPAC Extragalactic Database (NED), which is funded by the National Aeronautics and Space Administration and operated by the California Institute of Technology.

The DESI Legacy Imaging Surveys consist of three individual and complementary projects: the Dark Energy Camera Legacy Survey (DECaLS), the Beijing-Arizona Sky Survey (BASS), and the Mayall z-band Legacy Survey (MzLS). DECaLS, BASS and MzLS together include data obtained, respectively, at the Blanco telescope, Cerro Tololo Inter-American Observatory, NSF’s NOIRLab; the Bok telescope, Steward Observatory, University of Arizona; and the Mayall telescope, Kitt Peak National Observatory, NOIRLab. NOIRLab is operated by the Association of Universities for Research in Astronomy (AURA) under a cooperative agreement with the National Science Foundation. Pipeline processing and analyses of the data were supported by NOIRLab and the Lawrence Berkeley National Laboratory (LBNL). Legacy Surveys also uses data products from the Near-Earth Object Wide-field Infrared Survey Explorer (NEOWISE), a project of the Jet Propulsion Laboratory/California Institute of Technology, funded by the National Aeronautics and Space Administration. Legacy Surveys was supported by: the Director, Office of Science, Office of High Energy Physics of the U.S. Department of Energy; the National Energy Research Scientific Computing Center, a DOE Office of Science User Facility; the U.S. National Science Foundation, Division of Astronomical Sciences; the National Astronomical Observatories of China, the Chinese Academy of Sciences and the Chinese National Natural Science Foundation. LBNL is managed by the Regents of the University of California under contract to the U.S. Department of Energy. The complete acknowledgments can be found at \url{https://www.legacysurvey.org/acknowledgment/}.

We thank the referee for the insightful comments and suggestions that significantly improved our paper and its presentation.

\section*{Data Availability}

The radio imaging products presented here are made available with this paper at doi, including spectral index, and Stokes I, Q and U L-band cubes.

%%%%%%%%%%%%%%%%%%%% REFERENCES %%%%%%%%%%%%%%%%%%

% The best way to enter references is to use BibTeX:

\bibliographystyle{mnras}
\bibliography{ref} % if your bibtex file is called example.bib

% Don't change these lines
\bsp	% typesetting comment
\label{lastpage}
\end{document}